\documentclass[11pt,a4paper]{article}
\pdfoutput=1
\usepackage{jcappub}

\usepackage{bm}
\usepackage{amsfonts}
\usepackage{subfigure}
\usepackage{marginnote}

\newcommand{\be}{\begin{equation}}
\newcommand{\ee}{\end{equation}}
\newcommand{\bea}{\begin{eqnarray}}
\newcommand{\eea}{\end{eqnarray}}

\begin{document}



\title{Dark energy properties from large future galaxy surveys}

\author[a]{Tobias Basse}
\author[a]{Ole Eggers Bj{\ae}lde}
\author[b]{Jan Hamann}
\author[a,c]{Steen Hannestad}
\author[d]{Yvonne Y.Y. Wong}

\affiliation[a]{Department of Physics and Astronomy\\
 Aarhus University, DK-8000 Aarhus C, Denmark}

\affiliation[b]{Theory Division, Physics Department\\
 CERN, CH-1211 Geneva 23, Switzerland}

\affiliation[c]{Aarhus Institute of Advanced Studies,\\
 Aarhus University, DK-8000 Aarhus C, Denmark}

\affiliation[d]{School of Physics\\ The University of New South Wales, Sydney NSW 2052, Australia}

\emailAdd{basse@phys.au.dk, oeb@phys.au.dk, jan.hamann@cern.ch, sth@phys.au.dk, yvonne.y.wong@unsw.edu.au}

\abstract{
We perform a detailed forecast on how well a {\sc Euclid}-like survey will be able to constrain dark energy and neutrino parameters from a combination of its cosmic shear
power spectrum, galaxy power spectrum, and cluster mass function measurements.
We find that the combination of these three probes vastly improves the survey's potential to measure the time evolution of dark energy. In terms of a dark energy figure-of-merit defined as $(\sigma(w_{\mathrm p}) \sigma(w_a))^{-1}$,  we find a value of 690 for {\sc Euclid}-like data combined with {\sc Planck}-like measurements of the cosmic microwave background (CMB) anisotropies in a 10-dimensional cosmological parameter space, assuming a  $\Lambda$CDM fiducial cosmology.  For the more commonly used 7-parameter model, we find a figure-of-merit of 1900 for the same data combination.
We consider also the survey's potential to measure dark energy perturbations in models wherein the dark energy is parameterised as a fluid with a nonstandard non-adiabatic sound speed, and find that in an \emph{optimistic} scenario in which $w_0$ deviates by as much as is currently observationally allowed from $-1$, models with $\hat{c}_\mathrm{s}^2 = 10^{-6}$ and $\hat{c}_\mathrm{s}^2 = 1$ can be distinguished at more than $2\sigma$ significance. We emphasise that constraints on the dark energy sound speed from cluster measurements are strongly dependent on the modelling of the cluster mass function; significantly weaker sensitivities ensue if we modify our model to include fewer features of nonlinear dark energy clustering.
Finally, we find that the sum of neutrino masses can be measured with a $1 \sigma$ precision of 0.015~eV, even in complex cosmological models in which the dark energy equation of state varies with time. The $1 \sigma$ sensitivity to the effective number of relativistic species $N_{\rm eff}^{\rm ml}$ is approximately 0.03, meaning that  the small deviation  of 0.046 from 3 in the standard value of~$N_{\rm eff}^{\rm ml}$ due to non-instantaneous decoupling and finite temperature effects can be probed with $1\sigma$ precision for the first time.}
\maketitle

\section{Introduction}

The coming decade will see spectacular advances in the measurement of the large-scale structure distribution in the universe. Perhaps the most interesting of these measurements are the large-scale photometric surveys to be conducted by the Large Synoptic Survey Telescope (LSST)~\cite{Abell:2009aa} and the ESA {\sc Euclid} mission~\cite{Laureijs:2011mu}. Both projects will map the positions and measure the shapes of order a billion galaxies in a significant fraction of the current Hubble volume. This will in turn allow for a precision measurement of both the galaxy clustering and the cosmic shear power spectra, and likewise an impressively precise determination of the cosmological parameter values.   As an example, assuming a vanilla $\Lambda$CDM model extended with nonzero neutrino masses, a {\sc Euclid}-like survey will be able to measure the neutrino mass sum $\sum m_\nu$ at a precision of at least 0.03~eV~\cite{Carbone:2011by,Audren:2012vy} (most optimistically up to 0.01~eV~\cite{Hamann:2012fe}) when combined with measurements of the cosmic microwave background (CMB) anisotropies from the Planck mission~\cite{Planck:2006aa}.   Such a precision will see the absolute neutrino mass scale detected at high confidence
even if the true value of $\sum m_\nu$  should be the minimum compatible with current neutrino oscillation data, i.e., $\sum m_\nu \simeq 0.06$~eV~\cite{Hamann:2012fe}.

As yet unexplored in reference~\cite{Hamann:2012fe} is the role played by the cluster mass function in cosmological parameter inference.
Weak gravitational lensing measurements available to both the LSST and {\sc Euclid} will allow for the efficient detection and mass determination of galaxy clusters; {\sc Euclid}, for example, is expected to detect and accurately measure the masses of close to 100,000 clusters~\cite{Laureijs:2011mu}.
In this work, we continue the cosmological parameter sensitivity forecast begun in~\cite{Hamann:2012fe} by adding the cluster mass function inferred from a {\sc Euclid}-like cluster survey to the galaxy and the shear power spectra measurements already considered in reference~\cite{Hamann:2012fe}.
We also extend the study to dark energy models with a time-dependent  equation of state and/or a nonstandard non-adiabatic sound speed.
As in~\cite{Hamann:2012fe}  we shall adopt the survey specifications of the {\sc Euclid} mission in terms of the number of objects observed, the redshift range, and the sky coverage.  However, the analysis procedure can be easily adapted to other similar redshift surveys such as the LSST.

The paper is structured as follows.  We discuss first our dark energy parameterisation in section~\ref{sec:DEP}, before introducing in section~\ref{sec:euclidobs}
the cluster mass function as a cosmological observable.  In section~\ref{sec:errors} we examine some uncertainties likely to be encountered in a {\sc Euclid}-like
measurement of the cluster mass function, and discuss how we model and propagate these uncertainties in our forecast analysis.  Sections~\ref{sec:mockdatagen} and~\ref{sec:forecast} outline respectively our mock data generation and forecast procedures, while section~\ref{sec:results} contains our results.   We conclude in section~\ref{sec:conc}.

\section{Dark energy parameterisation}\label{sec:DEP}

Although dark energy is the most popular explanation for the apparent accelerated expansion of the universe,  there is as yet no consensus on its actual physical properties. For this reason, and for reasons of simplicity, dark energy is usually described as a fluid obeying the laws of general relativity. The homogeneous part of this fluid is responsible for
driving the expansion of the universe, and can be represented by an equation of state $w(\tau)=\bar{P}_Q(\tau)/\bar{\rho}_Q(\tau)$, where $\bar{P}_Q(\tau)$ and $\bar{\rho}_Q(\tau)$ denote the unperturbed dark energy pressure and energy density respectively, and $\tau$ is conformal time.  Except in the case of a cosmological constant, for which $w(\tau)$ is precisely the constant $-1$, dynamical dark energy models have in general equations of state that are functions of time. For this reason, we model dark energy equation of state  using the popular parameterisation~\cite{Chevallier:2000qy,Linder:2002et}
\begin{equation}\label{eq:lcp-param}
    w(\tau)=w_0+w_a[1-a(\tau)],
\end{equation}
where $w_0$ and $w_a$ are constants, and $a(\tau)$ denotes the scale factor. Note that this parameterisation should be regarded simply as a toy model that facilitates comparisons across different observational probes.  We make no pretence here that it actually captures the behaviour of any realistic dynamic dark energy model.
For an example of a forecast tailored specifically to scalar-field models of dark energy, see, e.g.,~\cite{Pavlov:2012fd}.

A general relativistic fluid evolving in an inhomogeneous spacetime will in general develop inhomogeneities of its own.
We express inhomogeneities in the dark energy density in terms of a density contrast $\delta_Q(\tau,\textbf{x})$ satisfying $\rho_Q(\tau,\textbf{x})=\bar\rho_Q(\tau)[1+\delta_Q(\tau,\textbf{x})]$, where $\rho_Q(\tau,\textbf{x})$ is the fully time- and space-dependent dark energy density.

The evolution of the density contrast $\delta_Q(\tau, \textbf{x})$ can be described by a set of (nonlinear) fluid equations coupled to the (nonlinear) Einstein equation.  For the nonlinear aspects of the formation of clusters we refer to section~\ref{sec:theory} and appendix~\ref{sec:appendix}. For the purpose of calculating the linear matter power spectrum we implement the linear evolution for the dark energy density contrast, described in the synchronous gauge
and in Fourier space by the equations of motion (see, e.g., \cite{DeDeo:2003te,Bean:2003fb,Weller:2003hw,Ballesteros:2010ks})
\begin{eqnarray}
\label{eq:eom}
&&{\dot \delta}_Q + (1+w)\left(\theta_Q+\frac{\dot{h}}{2}\right)+3(\hat{c}_\mathrm{s}^2-w)
{\cal H}
\delta_Q+9(1+w)(\hat{c}_\mathrm{s}^2-c_a^2)
{\cal H}^2
\frac{\theta_Q}{k^2} =0, \nonumber \\
&& {\dot {\theta}}_Q + (1-3\hat{c}_\mathrm{s}^2)
{\cal H}
\theta_Q-\frac{\hat{c}_\mathrm{s}^2k^2}{1+w}\delta_Q+k^2\sigma_Q=0,
\end{eqnarray}
where  $\delta_Q(\tau, k)$ now denotes the dark energy density contrast in Fourier $k$-space,
$\theta_Q(\tau,k)$ is the divergence of the dark energy velocity field, $h$ the metric perturbation,
${\cal H} \equiv \dot{a}/a$ the conformal Hubble parameter,  $\sigma_Q$ is the shear stress which we assume to be vanishing in this work, 
and ${\hat c}_\mathrm{s}^2\equiv\delta P_Q/\delta\rho_Q|_{\rm rest}$ and $c_a^2\equiv\dot{\bar{P}}_Q/\dot{\bar{\rho}}_Q$ are the non-adiabatic and adiabatic dark energy sound speeds respectively.  
Note that the non-adiabatic sound speed ${\hat c}_\mathrm{s}^2$ is defined as the ratio of the pressure perturbation $\delta P_Q$ to the energy density perturbation $\delta \rho_Q$ in the rest-frame of the dark energy fluid, while
the adiabatic sound speed $c_a^2$ is related to the homogeneous fluid equation of state via $\dot{w}=3(1+w)(w-c_a^2){\cal H}$.  We employ natural units throughout this work, i.e., $c=1$ denotes the speed of light.

\section{The cluster mass function as  a {\sc Euclid} observable}\label{sec:euclidobs}

Cluster surveys can be an excellent probe of dynamical dark energy because the abundance of the most massive gravitationally bound objects at any one time depends strongly on both the growth function of the matter perturbations and the late-time expansion history of the universe (see, e.g., \cite{Wang:2004pk,Takada:2007fq,Sefusatti:2006eu,Wang:2005vr,Lima:2005tt,Abramo:2009ne}).  The {\sc Euclid} mission will identify clusters in the photometric redshift survey  accompanied by a spectroscopic follow-up.
The same survey will also determine the masses of the detected clusters by way of weak gravitational lensing.

\subsection{Cluster mass function from theory}\label{sec:theory}

A simple quantification of the cluster distribution is the cluster mass function.   Denoted $\mathrm{d}n/\mathrm{d}M(M,z)$, the cluster mass function counts the number of clusters per comoving volume in a given mass interval $[M, M+\mathrm{d}M]$ as a function of redshift $z$.

For any given cosmological model, an accurate prediction of the corresponding cluster mass function necessitates the use of $N$-body/hydrodynamics simulations.  However, a number of fitting functions, calibrated against simulation results in the vanilla $\Lambda$CDM model framework, have been proposed in the literature (e.g.,~\cite{Sheth:1999mn,Jenkins:2000bv,Tinker:2008ff}).  In this work, we model
the cluster mass function after the Sheth-Tormen fitting function \cite{Sheth:1999mn}
\begin{equation}
	\frac{\mathrm{d}n_{\rm ST}}{\mathrm{d}M}(M,z)=-\sqrt{\frac{2a}{\pi}}A\left[1+\left(\frac{a\delta_c^2}{\sigma_\mathrm{m}^2}\right)^{-p}\right]\frac{\bar\rho_\mathrm{m}}{M^2}\frac{\delta_c}{\sigma_\mathrm{m}}\left(\frac{\mathrm{d}\log \sigma_\mathrm{m}}{\mathrm{d}\log M}-\frac{\mathrm{d}\log \delta_c}{\mathrm{d}\log M}\right)\exp\left[-a\frac{\delta_c^2}{2\sigma_\mathrm{m}^2}\right],
 \label{eq:sh}
\end{equation}
where the fitting parameters are $a=0.707$, $A=0.322184$, and $p=0.3$, and $\bar{\rho}_\mathrm{m}(z)$ is the mean matter density (it was shown in~\cite{Brandbyge:2010ge} that this function provides a very good fit also in models with non-zero neutrino mass).
The quantity $\sigma_\mathrm{m}^2(M,z)$ denotes the variance of the linear matter density field smoothed on a comoving length scale $X_{\rm sm} \equiv a^{-1} [3 M/(4 \pi \bar{\rho}_\mathrm{m})]^{1/3}$, and is computed from the linear matter power spectrum
 $P^\mathrm{lin}_\mathrm{m}(k,z)$ via
\begin{equation}
 \sigma_\mathrm{m}^2(M,z)\equiv\frac{1}{2\pi^2}\int^\infty_0 \mathrm{d}k\ k^2|W(k X_{\rm sm})|^2P^\mathrm{lin}_\mathrm{m}(k,z),
 \label{eq:sigma}
\end{equation}
where $W(x) = 3 \ (\sin x - x \cos x)/x^3$ is the Fourier transform of the spherical (spatial) top-hat filter function.   The linear power spectrum  $P^\mathrm{lin}_\mathrm{m}(k,z)$ can be obtained from a Boltzmann code such as {\sc Camb}~\cite{Lewis:1999bs}.

The quantity $\delta_c(M,z)$ is known as the linear threshold density of matter at the time of collapse.  Its value is established by tracking the full nonlinear collapse of a spherical top-hat over-density, noting the time $\tau_{\rm coll}$ the region collapses to an infinitely dense point, and then computing from {\it linear} perturbation theory the {\it linear} density contrast at  $\tau= \tau_{\rm coll}$.  In many applications it suffices to take the constant value $\delta_c = 1.68$.  In dark energy cosmologies, however, this may not be a very good approximation (see, e.g., \cite{Mota:2004pa,Basse:2010qp}). Here, we estimate $\delta_c(M,z)$ as described immediately above, and track the spherical collapse of a top-hat
overdensity by solving the equations%
\begin{eqnarray}
\label{eq:ddotr}
&& \frac{\ddot{X}}{X} + {\cal H} \frac{\dot{X}}{X} = - \frac{4 \pi G}{3} a^2 [ \overline{\rho}_\mathrm{m} \delta_\mathrm{m}+
 \overline{\rho}_Q(1 + 3 \hat{c}_\mathrm{s}^2) \delta_Q], \\
&& \delta_\mathrm{m} (\tau) = [1+ \delta_\mathrm{m} (\tau_i)] \left[ \frac{X(\tau_i)}{X (\tau)}  \right]^3 -1,
\label{eq:deltaspherical}
\end{eqnarray}
where $X(\tau)$ is the comoving radius of the top-hat, $\delta_\mathrm{m}(\tau)$  and $\delta_Q(\tau)$ the matter and the dark energy density contrasts respectively in the top-hat region, and $\tau_i$  is a reference initial time.   Note that equation~(\ref{eq:deltaspherical}) follows from conservation of the total mass of nonrelativistic matter~$M_\mathrm{m}$ in the top-hat region.
For more detailed discussions of the spherical collapse model, we direct the reader to references~\cite{Gunn:1972sv,Wang:1998gt,Cooray:2002dia}.

The presence of the $\overline{\rho}_Q(1 + 3 \hat{c}_\mathrm{s}^2) \delta_Q$ term on the right hand side of equation~(\ref{eq:ddotr}) indicates that
the dark energy component also participates in the collapse, especially when the initial dimension $X(\tau_i)$ of the top-hat matter overdensity exceeds the comoving Jeans length associated with the fluid's non-adiabatic sound speed~\cite{Abramo:2007iu,Creminelli:2009mu,Basse:2010qp,Basse:2012wd}.
The resulting linear threshold density $\delta_c$ therefore exhibits  generically a dependence on the mass of the collapsing region, in addition to the usual $z$-dependence.
However, tracking the nonlinear evolution of the dark energy density contrast $\delta_Q$ is in general nontrivial because the spherical top-hat region is well-defined strictly only in the $\hat{c}_\mathrm{s}^2=0$ and the $\hat{c}_\mathrm{s}^2 \to \infty$ limits, where, supplemented with~\cite{Creminelli:2009mu,Basse:2010qp}
\begin{eqnarray}
&& \overline{\rho}_Q(1 + 3 \hat{c}_\mathrm{s}^2) \delta_Q \to 0,  \qquad \qquad \qquad  \qquad \hat{c}_\mathrm{s}^2 \to \infty,  \\
&&\dot{\rho}_Q + 3 \left( {\cal H} + \frac{\dot{X}}{X} \right) (\rho_Q + \bar{P}_Q) = 0, \qquad  \hat{c}_\mathrm{s}^2 =0,
\label{eq:cs0conservation}
\end{eqnarray}
the collapse equation~(\ref{eq:ddotr}) can be solved exactly.   Extending the application of equation~(\ref{eq:ddotr})
to the intermediate regime necessitates additional assumptions, which do not however render the system any less intractable~\cite{Basse:2010qp,Basse:2012wd}.  For this reason, we shall resort to modelling the mass-dependence of $\delta_c(M,z)$ by interpolating between the two known limits using a hyperbolic tangent function, where the location of the kink at each redshift is adjusted to reflect the Jeans mass corresponding  to the given~$\hat{c}_\mathrm{s}^2$.
 See appendix~\ref{sec:appendix} for details.

Finally, the virial radius $R_{\rm vir}$ and the virial mass $M_{\rm vir}$ can likewise be computed from equations~(\ref{eq:ddotr}) to~(\ref{eq:cs0conservation}) in the two limits of $\hat{c}_\mathrm{s}^2$.  Here,  $R_{\rm vir} \equiv a X(\tau_{\rm vir})$ is defined as the physical radius of the top-hat region and $M_{\rm vir}$ the total mass contained therein at virialisation, where virialisation is taken to mean the moment at which the virial theorem is satisfied by the collapsing region,~$\tau_{\rm vir}$.   The virial mass
$M_{\rm vir} = M_\mathrm{m}+ M_Q(\tau_{\rm vir})$ counts both contributions from nonrelativistic matter  $M_\mathrm{m}$ and from the clustered dark energy
$M_Q \equiv (4 \pi/3) \bar{\rho}_Q \delta_Q (aX)^3$, and it is $M_{\rm vir}$, not~$M_\mathrm{m}$, that we identify with the cluster mass $M$ throughout this work.  Further details can be found in
appendix~\ref{sec:appendix}.  The virial radius will be used in section~\ref{sec:thr} to determine the cluster mass detection threshold.

\subsection{The observable}

The actual observable quantity is the number of clusters $N_{i,j(i)}$ in the redshift bin~$i$ and the (redshift-dependent) mass bin~$j(i)$, defined as
\begin{equation}
N_{i,j(i)} = \Delta\Omega\int_0^\infty   \mathrm{d}z \int_0^\infty \mathrm{d}M \
\frac{\mathrm{d}^2V}{\mathrm{d}\Omega \mathrm{d}z}\left(z\right)W_{i,j(i)} (M,z)\frac{\mathrm{d}n_{\rm ST}}{\mathrm{d}M} (M,z),
 \label{eq:Nalbe}
\end{equation}
where $\Delta \Omega$ is the solid angle covered by the survey (taken to be $15,000\,{\rm deg}^2$ following~\cite{Laureijs:2011mu}), $\mathrm{d}^2V/(\mathrm{d}\Omega \mathrm{d}z)(z)$  the comoving volume element at redshift~$z$, and $W_{i,j(i)}(M,z)$
is the window function defining the redshift and mass bin.  Note that the window functions are in general not sharp in $z$- and $M$-space
because of uncertainties in the redshift and the mass determinations (see sections~\ref{sec:sigmaz} and~\ref{sec:masserr}).
We defer the discussion of our binning scheme to section~\ref{sec:binning}.

\section{Measurement errors}\label{sec:errors}

\subsection{Redshift uncertainty}\label{sec:sigmaz}

The photometric survey will measure redshifts with an estimated scatter of \mbox{$\sigma (z) \sim 0.03 (1+z)$} and almost no bias~\cite{Laureijs:2011mu}.  Nonetheless, because the  detected clusters will be subject to a follow-up spectroscopic study, the effective uncertainty in the redshift determination {\it per se} can be taken as negligible.
Additional redshift errors may arise from the  peculiar velocities of the clusters, where velocities up to $\sim 1000 \, {\rm km} \, {\rm s}^{-1}$ may lead to an error of $\delta z \sim 0.003$.  However, as we shall see in section~\ref{sec:binning}, even the narrowest redshift bins adopted in our analysis typically have widths of order $\Delta z \sim 0.03$, i.e., a factor of ten larger than the peculiar velocity uncertainty.   We therefore treat the cluster redshift as infinitely well-determined, and approximate the window function of the redshift bin~$i$ as
\begin{equation}
\label{eq:zwindow}
\theta(z-z_{\min,i}) \theta(z_{\max, i}-z),
\end{equation}
where $z_{\min,i}$ and $z_{\max,i}$ denote, respectively, the lower and upper boundaries of the bin, and $\theta$ is the Heaviside step function.

\subsection{Uncertainty in the weak lensing mass determination}\label{sec:masserr}

The mass of a cluster determined through weak lensing, $M_{\rm obs}$, is subject to scatter and bias with respect to the true mass of the cluster $M$~\cite{Bahe:2011cb,Becker:2010xj}.
For a mass determination algorithm that treats clusters as spherical objects, the triaxiality of realistic cluster density profiles, for example, could cause the cluster mass to be over- or underestimated depending on the orientation of the major axis in relation to the line-of-sight.
Additional biases are incurred if the true density profile deviates from the assumed one.

In this work, we assume that the bias can be controlled to the required level of accuracy, and model only the scatter in the mass determination using a log-normal distribution \cite{Bahe:2011cb,Becker:2010xj},
\begin{equation}
P\left(M_{\rm obs}|M\right) = \frac{1}{M_{\rm obs} \sqrt{2\pi\sigma^2}}\exp\left[-\frac{\left(\ln M_{\rm obs}-\mu\right)^2}{2\sigma^2}\right],
\label{eq:log-normal}
\end{equation}
whose mean is given by $\exp(\mu + \sigma^2/2)$.
Here, $P\left(M_{\rm obs}|M\right)$ denotes the probability that a cluster with true mass $M$ is mistakenly determined to have a mass $M_{\rm obs}$ by the survey.
Since we assume an unbiased mass determination, it follows that the mean of the distribution must match the true mass $M$ and subsequently $\mu = \ln M - \sigma^2/2$.  We use $\sigma = 0.6$ to model the mass scatter~\cite{Bahe:2011cb}.

The distribution~(\ref{eq:log-normal})  can be integrated over $M_{\rm obs}$ in the interval $[M_{\min,j(i)},M_{\max,j(i)}]$  in order to determine the probability that a cluster of true mass $M$ in the redshift bin~$i$ will be determined to lie in the mass bin $j(i)$.  Combing the resulting integral with the redshift window function $V_i(z)$ from equation~(\ref{eq:zwindow}), we obtain the window function for the redshift and mass bin~$\{i, j(i)\}$,
\begin{equation}
W_{i,j(i)} (M,z) = \theta(z-z_{\min,i}) \theta(z_{\max, i}-z)   \int_{M_{\min,j(i)}}^{M_{\max,j(i)}} \mathrm{d} M_{\rm obs} \
P\left(M_{\rm obs}|M\right) \theta(M_{\rm obs}-M_{\rm thr}(z)),
\label{eq:probinbin}
\end{equation}
where $M_{\rm thr}(z)$ is the mass detection threshold, to be discussed in section~\ref{sec:thr}.

\section{Mock data generation}\label{sec:mockdatagen}

The observable quantity in a cluster survey is the number of clusters $N_{i,j(i)}$ in the redshift bin~$i$ and mass bin~$j(i)$.  Thus for any given fiducial cosmology and survey specifications, one may compute the fiducial cluster numbers $N^{\rm fid}_{i,j(i)}$ as per equation~(\ref{eq:Nalbe}), and then create a mock data set $\hat{N}_{i,j(i)}$ by assuming $\hat{N}_{i,j(i)}$ to be a stochastic variable that follows a Poisson distribution with parameter  $N^{\rm fid}_{i,j(i)}$.   An ensemble of realisations may be generated by repeating the procedure multiple times, and parameter inference performed on each mock realisation in order to assess the performance of a survey.

This is clearly a very lengthy process.  However, as shown in reference~\cite{Perotto:2006rj},  for the sole purpose of establishing a survey's sensitivity to cosmological parameters, it suffices to use only one mock data set in which the data points are set to be equal to the predictions of the fiducial model, i.e.,  $\hat{N}_{i,j(i)} = N^{\rm fid}_{i,j(i)}$.  This much simplified procedure correctly reproduces the survey sensitivities in the limit of infinitely many random realisations, and is the procedure we adopt in our analysis.

In the following we describe in some detail the survey specifications that go into the computation of the fiducial $N^{\rm fid}_{i,j(i)}$: the mass detection threshold, our redshift and mass binning scheme, and the survey completeness and efficiency.  In section~\ref{sec:datasets} we summarise the mock data sets to be used in our parameter sensitivity forecast.

\subsection{Mass detection threshold}\label{sec:thr}

We model the redshift-dependent mass detection threshold $M_{\rm thr} (z)$ following the approach of references~\cite{Wang:2004pk,Hamana:2003ts}.
A cluster of mass $M$ at redshift~$z$ produces a shear signal $\kappa_G(M,z)$, where
\begin{equation}
\kappa_G (M,z)= \alpha (M,z)
\frac{M/[\pi R_\mathrm{s}^2(M,z)]}{ \Sigma_\mathrm{cr}(z)}.
\label{eq:kappa}
\end{equation}
Here, assuming a truncated Navarro-Frenk-White (NFW) density profile~\cite{Navarro:1995iw},  $R_\mathrm{s} (M,z) = R_\mathrm{vir}(M,z)/c_\mathrm{nfw}$, where $R_{\rm vir}$ is the cluster's virial radius computed according to the spherical collapse model outlined in section~\ref{sec:theory} and appendix~\ref{sec:appendix},
and $c_\mathrm{nfw}$ is the halo concentration parameter determined from $N$-body simulations. We use $c_\mathrm{nfw} = 5$ following~\cite{Bahe:2011cb}.
The factor $\alpha(M,z)$ is computed from smoothing the (projected) NFW profile using a Gaussian filter of angular smoothing scale $\theta_G$, i.e.,
\begin{equation}
\alpha(M,z)= \frac{\int_0^\infty{\mathrm{d}x\left(x/x_G^2\right)\exp\left(-x^2/x_G^2\right)f_\mathrm{nfw}\left(x\right)}}{\ln\left(1+c_\mathrm{nfw}\right)-c_\mathrm{nfw}/\left(1+c_\mathrm{nfw}\right)},
\label{eq:alpha}
\end{equation}
where $x \equiv \theta/\theta_\mathrm{s}$ and  $x_G \equiv \theta_G/\theta_\mathrm{s}$,
with $\theta_\mathrm{s} (M,z)= R_\mathrm{s}(M,z)/d_A(z)$ and $d_A\left(z\right)$ the angular diameter distance to the cluster.
The projected NFW profile is encoded in the dimensionless surface density profile $f_\mathrm{nfw}\left(x\right)$, which can be found in equation~(7) of reference~\cite{Hamana:2003ts}. In our analysis we use an angular smoothing scale of $\theta_G = 1$~arcmin.

The mean critical surface mass density $\Sigma_\mathrm{cr}(z)$ is, assuming a flat spatial geometry, given by the expression
\begin{equation}
\Sigma_\mathrm{cr}^{-1}(z) = \frac{4\pi G}{\left(1+z\right)} \  n_{\rm bg}^{-1}  \int_{z}^\infty \mathrm{d}z' \; \mathrm{d}n/\mathrm{d}z' \,  \chi\left(z\right)  \left[1-\chi\left(z\right)/\chi\left(z'\right)\right],
\label{eq:Sigma}
\end{equation}
where $\chi(z')$ denotes the comoving radial distance to the redshift~$z'$, and $(\mathrm{d}n/\mathrm{d}z') \mathrm{d}z'$ is the number density of source galaxies per steradian at redshift $(z', z'+\mathrm{d}z')$, normalised such that $n_\mathrm{bg} = \int_0^\infty \mathrm{d}z'\; \mathrm{d}n/\mathrm{d}z'$ gives the source galaxy surface density.
As in~\cite{Hamann:2012fe}, we assume a galaxy redshift distribution of the form
\begin{equation}
\frac{\mathrm{d}n}{\mathrm{d}z} \propto \left(\frac{z}{z_0}\right)^2 \exp\left[-\left(\frac{z}{z_0}\right)^\beta\right],
\label{eq:dndz}
\end{equation}
where for a {\sc Euclid}-like survey we choose  $\beta = 1$, $z_0 = 0.3$, and a source galaxy surface density of $n_{\rm bg} =30\,\mathrm{arcmin}^{-2}$~\cite{Laureijs:2011mu}.

In order for a cluster to be considered detected, its shear signal $\kappa_G$ must exceed the ``noise'' of the survey $\sigma_{\rm noise}$ by a predetermined amount.
Shear detection is limited firstly by the intrinsic ellipticity of the background galaxies, and secondly by the number of galaxy images lensed by the cluster that fall within the smoothing aperture.  Thus, the noise term may be estimated as~\cite{VanWaerbeke:1999wv}
\begin{equation}
\sigma^2_\mathrm{noise} = \frac{\sigma^2_\epsilon}{4\pi \theta_G^2 n_\mathrm{bg}},
\label{eq:lnoise}
\end{equation}
where $\sigma_\epsilon$ denotes the total mean dispersion of the galaxy intrinsic ellipticity, and we use $\sigma_\epsilon = 0.35$ in our analysis.

Defining a signal-to-noise ratio of $S/N=3$ to be our detection threshold~\cite{Laureijs:2011mu},  the expression
\begin{equation}
S/N = \frac{\kappa_G(M_{\rm thr}(z), z) }{\sigma_\mathrm{noise}}
\end{equation}
can now be solved for the mass detection threshold $M_{\rm thr}(z)$.  This sets a lower limit on the cluster mass detectable by lensing at a given redshift $z$.

\subsection{Redshift and mass binning}\label{sec:binning}

We consider a survey that observes clusters in the redshift range $[z_{\rm low},z_{\rm high}]$.  We subdivide this range into $N_z$ bins in such a way so as  to maintain the same number of clusters in all bins~$i$ in the fiducial cosmology.  The resulting bin boundaries $z_{\min, i}$ and $z_{\max,i}$ then define the redshift window functions~(\ref{eq:zwindow}).  Clusters in each redshift bin are further subdivided according to their observed masses into $N_\mathrm{m}$ mass bins labelled by $j(i)$, again with the enforcement that the number of clusters $N_{i,j(i)}$ should be similar in all bins.

 An immediate consequence of such a binning scheme is a variation of the mass bin boundaries $M_{\min, j(i)}$ and $M_{\max, j(i)}$  with redshift because of  (i) the $z$-dependence of the mass detection threshold $M_{\rm thr}(z)$, and (ii) the rarity of high-mass clusters at high redshifts.   For the latter point, we impose in practice an absolute high-mass cut-off of $M_{\rm high} = 10^{16}~M_\odot$, i.e., the upper limit  of the last mass bin, $M_{\max, N_\mathrm{m}(i)}$, is always equal to $M_{\rm high}$ at all redshifts.
 This number also sets the high-redshift cut-off $z_{\rm high}$, which is defined to be the redshift at which the mass detection threshold $M_{\rm thr}(z)$ exceeds $M_{\rm high}$.
The lower cut-off is set at $z_{\rm low} = 0.01$, since the survey contains a negligible number of clusters below this redshift because of the small volume and
a large detection threshold.

\begin{figure}[t]
\center
\includegraphics[height=.48\textwidth,angle=270]{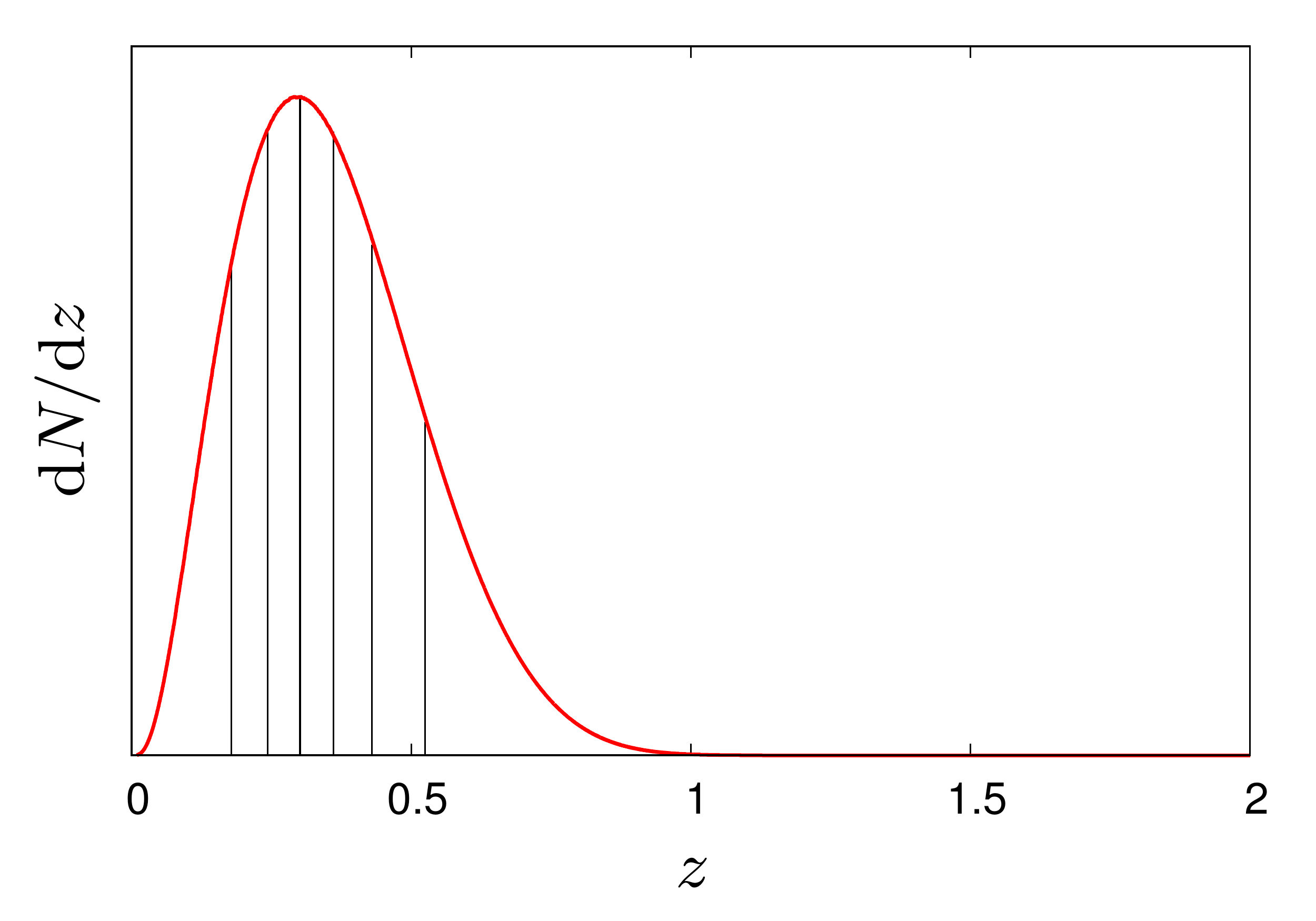}
\includegraphics[height=.48\textwidth,angle=270]{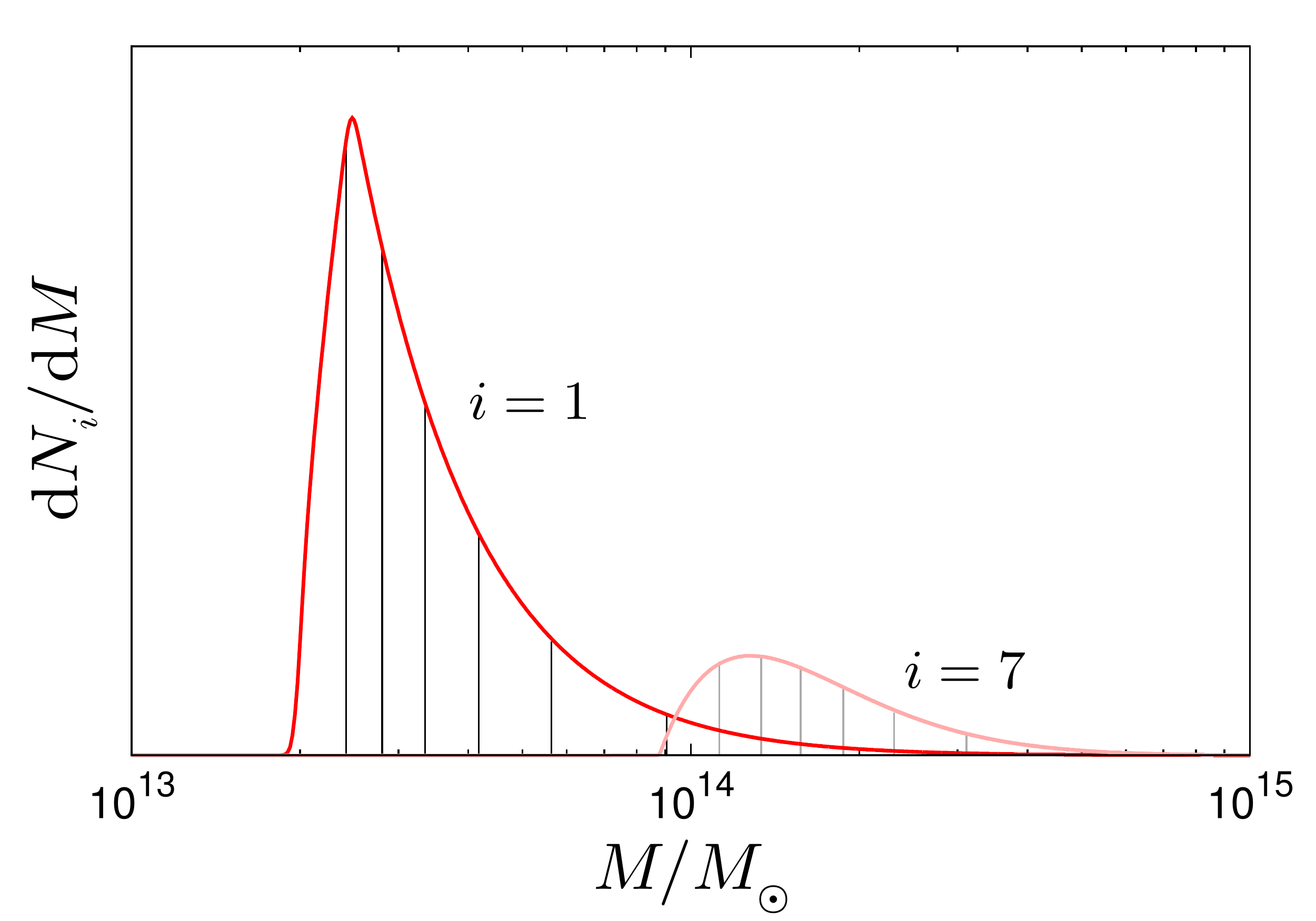} \\
\caption{
The left panel shows the division into of the observed number of clusters into 7 redshift bins while keeping the cluster count common for all bins.
The right panel shows the subsequent division of redshift bins $i=1,7$ into 7 mass bins, again with the stipulation that all mass bins contain the same number of clusters.\label{fig:illbin}
}
\end{figure}

Figure~\ref{fig:illbin} illustrates the division of the observed number of clusters into redshift and mass bins in the case $N_z=N_\mathrm{m}=7$. The left panel shows the first division in redshift, while the right panel shows the subsequent division of redshift bins $i=1,7$ into mass bins.

\subsection{Completeness and efficiency}\label{sec:efficiency}

The completeness $f_c$ of a cluster survey is defined as the fraction of clusters actually detected as peaks by the cluster finding algorithm, while the efficiency $f_e$ is the fraction of detected peaks that correspond to real clusters.   In general these quantities can be established precisely only with the help of mock cluster catalogues generated from $N$-body simulations (see, e.g.,~\cite{White:2001gs,Feroz:2008rm,Pace:2007ac,Hennawi:2004ai,Lima:2005tt,Hamana:2003ts}).
Here, we adopt the same simplistic approach taken in reference~\cite{Wang:2005vr}, and assume both $f_c$ and $f_e$ to be mass- and redshift-independent. 

The effect of a survey completeness and efficiency not equal to unity can be estimated from simple considerations.   The observable $N$ computed from theory as introduced in equation~(\ref{eq:Nalbe}) is the number of \emph{detectable} clusters in a survey.  However, the cluster finding algorithm will typically detect only a fraction of these, say, $N_{\rm peak}$ peaks, of which $N_{\rm false}=(1-f_e) N_{\rm peak}$ do not correspond to real clusters at all.   Thus, the number of \emph{detected} clusters to the \emph{detectable} clusters are related by
\begin{equation}
N = \frac{1}{f_c}\left(N_{\rm peak}-N_{\rm false}\right),
\end{equation}
where $N_{\rm peak} = f_c N/f_e$, and  $N_{\rm false} = \left(1/f_e-1\right)f_cN$.  Since both $N_{\rm peak}$ and $N_{\rm false}$ follow Poisson statistics, i.e., with variances $\sigma^2(N_{\rm peak})\sim N_{\rm peak}$ and  $\sigma^2(N_{\rm false}) \sim N_{\rm false}$, the variance of $N$ in a realistic survey can be estimated to be
\begin{align}
\nonumber \sigma^2\left(N\right) &= \frac{1}{f_c^2}\left[\sigma^2\left(N_{\rm peak}\right)+\sigma^2\left(N_{\rm false}\right)\right] 
\\&= N\left[1/f_e+\left(1/f_e-1\right)\right]/f_c.
\end{align}
From the second equality we see that $f_c \neq f_e \neq 1$ simply amounts to increasing the uncertainty on each individual data point by a factor $\sqrt{[1/f_e+(1/f_e-1)]/f_c}$, which can be incorporated into the forecast analysis at the level of the likelihood function. We shall return to this point in section~\ref{sec:likelihood} where we discuss explicitly the construction of the likelihood function.  Suffice it to say  for now that  we adopt the values $f_c=0.70$ and $f_e=0.75$, which may be reasonably expected for the LSST~\cite{Wang:2005vr}, and
which are likely very conservative when applied to {\sc Euclid} because of its much narrower point spread function.

\subsection{Synthetic data sets}\label{sec:datasets}

We summarise here the mock {\sc Euclid}-like data sets we generate and use in our parameter sensitivity forecast.

\begin{itemize}
\item A cluster data set in the redshift range $z \in [0.01, z_{\rm high}]$, and the mass range $M \in [ M_{\rm thr}(z), 10^{16}~M_\odot]$,
where $M_{\rm thr}(z)$ denotes the redshift-dependent mass detection threshold as described in section~\ref{sec:thr}, and $z_{\rm high}$ is defined as the redshift at which
$M_{\rm thr}(z)$ exceeds $10^{16}~M_\odot$ as discussed in section~\ref{sec:binning}.  We slice the redshift- and the mass-space into $N_z$ and $N_\mathrm{m}$ bins respectively
according to the scheme detailed in section~\ref{sec:binning}.

\item Mock data from a {\sc Planck}-like CMB measurement, generated according to the procedure of~\cite{Perotto:2006rj}.   Note that although we do not use real {\sc Planck} data~\cite{Ade:2013xsa}, only synthetic CMB data of comparable constraining power, we shall continue to refer to this synthetic data set as ``{\sc Planck} data'' when discussing parameter constraints.

\item We use also the cosmic shear auto-correlation power spectrum, the galaxy  clustering auto-spectrum, and the shear-galaxy cross-correlation power spectrum that will be derived from a {\sc Euclid}-like photometric survey.   The procedure for generating these mock data sets has already been described detail in~\cite{Hamann:2012fe}, and is recapitulated here for completeness.

\begin{itemize}

\item The cosmic shear auto-spectrum is $C_{\ell,ij}^{\rm ss}$, where the multipole $\ell$ runs from 2 to $\ell_{\rm max}^{\rm s}=2000$ independently of redshift.
The indices $i,j \in [1,N_{\rm s}]$ label the redshift bin, where the redshift slicing is such that all bins contain similar numbers of source galaxies and so suffer the same amount of shot noise.  We use $N_{\rm s}=2$; introducing more redshift bins does not significantly improve the parameter sensitivities~\cite{Hamann:2012fe}.

\item The galaxy auto-spectrum $C_{\ell,ij}^{\rm gg}$ comprises multipole moments running from $\ell=2$ to $\ell^{{\rm g},i}_{\rm max}$ in redshift bins $i,j \in [1,N_{\rm g}]$,
where the choice of $N_{\rm g}=11$ exhausts to a large extent the information extractable from $C_{\ell,ij}^{\rm gg}$~\cite{Hamann:2012fe}.
The redshift slicing is again designed to maintain the same number of source galaxies across all redshift bins.  In contrast to the cosmic shear auto-spectrum, 
we implement here also a redshift-bin-dependent maximum multipole $\ell^{{\rm g},i}_{\rm max}$ so as to eliminate those (redshift-dependent) scales on which nonlinear scale-dependent galaxy bias becomes important.  The linear galaxy bias is however always assumed to be exactly known.

\item Finally, the shear-galaxy cross-spectrum $C_{\ell,ij}^{\rm sg}$ in the shear redshift bin $i \in [1,N_{\rm s}]$ and galaxy redshift bin $j \in [1,N_{\rm g}]$ 
runs from $\ell=2$ to $\ell^{{\rm g},j}_{\rm max}$ determined by the galaxy redshift binning.

\end{itemize}

\end{itemize}

\section{Forecasting}\label{sec:forecast}

We now describe our parameter sensitivity forecast for a  {\sc Euclid}-like photometric survey including a measurement of the  cluster mass function.
The forecast is based on the construction of a likelihood function for the mock data, whereby the survey's sensitivities to cosmological parameters can be explored using
Bayesian inference techniques.

\subsection{Model parameter space\label{sec:fiducial}}

Reference~\cite{Hamann:2012fe} considered a 7-parameter space  spanned by the physical baryon density $\omega_\mathrm{b}$, the physical dark matter (cold dark matter and massive neutrinos) density $\omega_{\rm dm}$, the dimensionless Hubble parameter $h$,
the amplitude and spectral index of the primordial scalar fluctuations $A_{\rm s}$  and $n_{\rm s}$, the reionisation redshift  $z_{\rm re}$, and the neutrino density fraction \mbox{$f_\nu = \omega_\nu/\omega_{\rm dm}$}, with \mbox{$\omega_\nu = \sum m_\nu/(94.1 \; {\rm eV})$}.
  In the present analysis we extend this model parameter space  to include also the possibility of a non-standard radiation content, quantified by the effective number of massless neutrinos $N^{\rm ml}_{\rm eff}$, as well as three dynamical dark energy parameters \mbox{$\Theta^Q \equiv (w_0, w_a,\hat{c}_\mathrm{s}^2)$}, taking the total number of free parameters to eleven:
\begin{equation}
\label{eq:model}
\Theta^{(11)} \equiv (\Theta^{(8)}, \Theta^Q)
 \equiv \left( (\omega_{\rm b},\omega_{\rm dm},h,A_{\rm s},n_{\rm s},z_{\rm re},f_\nu,N^{\rm ml}_{\rm eff} ), (w_0,w_a,\hat{c}_\mathrm{s}^2 ) \right).
\end{equation}
As in~\cite{Hamann:2012fe} we assume only one massive neutrino state, so that $N^{\rm ml}_{\rm eff}=2.046+\Delta N$, where $\Delta N$ parameterises any non-standard physics that may induce a non-standard radiation content.
Note that the $0.046$ contribution in $N_{\rm eff}^{\rm ml}$ comes from non-instantaneous neutrino decoupling and finite temperature QED effects,
and should in principle be shared between {\it both} the massless and the massive neutrino states.  However, in practice, the precise treatment of this small correction has no measurable effect on our parameter forecast.
Lastly,  we remark that, parameterised as such, $\Delta N$ can run from the lowest value of $-2.046$ to anything positive.  Many popular models with  non-standard radiation contents associate a positive $\Delta N$ with additional relativistic particle species such as, e.g., sterile neutrinos \cite{Barbieri:1990vx,Kainulainen:1990ds}.
A negative $\Delta N$ can however arise in, e.g., models with extremely low reheating temperatures~\cite{Kawasaki:2000en,Hannestad:2004px}.

For the non-dark energy part of the parameter space, our fiducial model is defined by  the parameter values
\begin{equation}
\Theta^{(8)}_{\rm fid} = (0.022, 0.1126228, 0.7, 2.1 \times 10^{-9}, 0.96, 11, 0.00553, 2.046).
\end{equation}
For the dark energy sector, we begin with the fiducial values $\Theta^Q_{\rm fid}= (-1,0,\infty)$ corresponding to dark energy in the form of a cosmological constant.
The first part of our analysis (up to and including section~\ref{sec:FoM}) will also be performed with the dark energy sound speed fixed at $c_\mathrm{s}^2 = \infty$, i.e., homogeneous dark energy.  We shall return to dark energy density perturbations in section \ref{sec:perturb}, and study the constraints on the dark energy sound speed under
a variety of assumptions for the fiducial dark energy parameter values $\Theta_{\rm fid}^Q$.

\subsection{Likelihood function}\label{sec:likelihood}

Given a theoretical prediction ${N}_\mathrm{th}$ for the observable number of clusters in a specific redshift and mass bin, the probability of actually observing ${N}_\mathrm{obs}$ clusters follows a Poisson distribution of $N_{\rm obs}$ degrees of freedom.  However, the imperfect completeness and efficiency of the survey necessitate that we rescale uncertainty on each data point by an amount $f^{-1} \equiv \sqrt{[1/f_e+(1/f_e-1)]/f_c}$ (see section~\ref{sec:efficiency}).  We accomplish this by defining an effective number of observed clusters $\widetilde{N}_{\rm obs} \equiv f^2 N_{\rm obs}$, and likewise an effective theoretical prediction $\widetilde{N}_{\rm th} \equiv f^2 N_{\rm th}$.  The effective probability distribution is then
\begin{equation}
\label{eq:poison}
\mathcal{L}_\mathrm{P}\left(\widetilde{N}_\mathrm{obs}|\widetilde{N}_\mathrm{th}\right) = \frac{\widetilde{N}_\mathrm{th}^{\widetilde{N}_\mathrm{obs}}}{\widetilde{N}_\mathrm{obs}!} \, \exp \left[ -\widetilde{N}_\mathrm{th} \right].
\end{equation}
In a real survey, the effective observed number of clusters $\widetilde{N}_\mathrm{obs}$ in any one bin is necessarily an integer so that equation~(\ref{eq:poison}) applies directly.  In our forecast, however, $\widetilde{N}_\mathrm{obs}$ corresponds to the theoretical expectation value of the fiducial model which generally does not evaluate to an integer.  To circumvent this inconvenience, we generalise  the likelihood function~(\ref{eq:poison}) by linearly interpolating the logarithm of the discrete distribution $\mathcal{L}_\mathrm{P}$
 in the interval $[ \mathrm{floor} ( \widetilde{N}_\mathrm{obs}),  \mathrm{ceiling} ( \widetilde{N}_\mathrm{obs}) ]$, i.e.,
\begin{equation}
\begin{aligned}
\ln {\mathcal{L}}( \widetilde{N}_\mathrm{obs}|\widetilde{N}_\mathrm{th})
 \equiv & \left(1  + \mathrm{floor} (\widetilde{N}_\mathrm{obs}) - \widetilde{N}_\mathrm{obs} \right)  \ln \mathcal{L}_\mathrm{P}\left(\mathrm{floor} (\widetilde{N}_\mathrm{obs} )|\widetilde{N}_\mathrm{th} \right)  \\
& + \left( \widetilde{N}_\mathrm{obs} - \mathrm{floor} ( \widetilde{N}_\mathrm{obs} ) \right) \ln \mathcal{L}_\mathrm{P} \left( \mathrm{ceiling} (\widetilde{N}_\mathrm{obs})|\widetilde{N}_\mathrm{th} \right).
\end{aligned}
\end{equation}
The total cluster log-likelihood function is then obtained straightforwardly by summing $\ln {\mathcal{L}}$ over all redshift and mass bins.

\section{Results}\label{sec:results}

\subsection{Impact of the number of bins}

\begin{figure}[t]
\center
\includegraphics[height=.96\textwidth,angle=270]{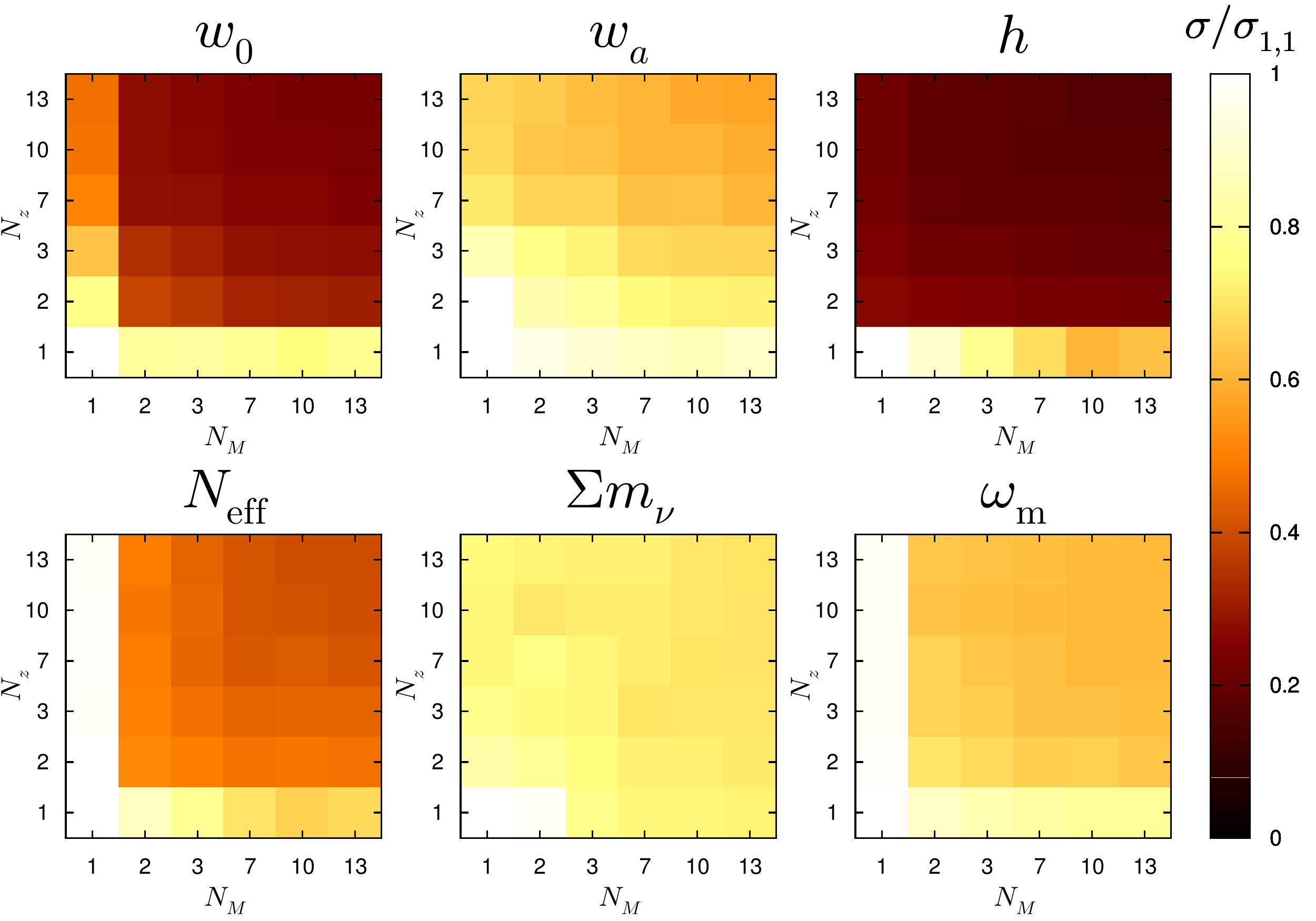}
\caption{Dependence of the posterior standard deviations, $\sigma$, for selected cosmological parameters for CMB+clusters on the number of redshift bins~$N_z$ and mass bins~$N_\mathrm{m}$. All numbers have been normalised to the corresponding $N_z=N_\mathrm{m} = 1$ result.}\label{fig:binning}
\end{figure}

We examine first how the parameter sensitivities of the cluster survey depend on the number of redshift and mass bins used. Figure~\ref{fig:binning} shows the posterior standard deviations for a range of cosmological parameters derived from a combination of synthetic  {\sc Planck} and cluster data as functions of $N_z$ and $N_\mathrm{m}$, normalised to the corresponding $N_z=N_\mathrm{m}=1$ result.

A general trend is immediately clear: for the parameters $w_0$ and $h$,
while increasing the number of mass bins results in moderate gain, it is the number of redshift bins used that contributes mostly to improving the parameter sensitivities.  For example, in the case of a fixed $N_\mathrm{m}=1$, the number of redshift bins needs to be increased to two or three in order for the sensitivities to improve as much or more than what can be gained for a fixed $N_z=1$ by splitting the data into ten or more redshift bins.  For $N_{\rm eff}^{\rm ml}$ and~$\omega_\mathrm{m}$ improvements in sensitivity are absent when no mass binning is used. From $N_\mathrm{m} = 2$ and beyond decent improvements are found in both directions. This can be traced to the fact that these parameters are primarily responsible for the shape and the overall normalisation of the cluster mass function, less so the redshift dependence.  See section~\ref{sec:expansionvspk}. For $w_a$ and $\sum m_\nu$ the sensitivity increases roughly equally with incresing $N_z$ and $N_\mathrm{m}$.

\begin{figure}[t]
\center
\includegraphics[height=.48\textwidth,angle=270]{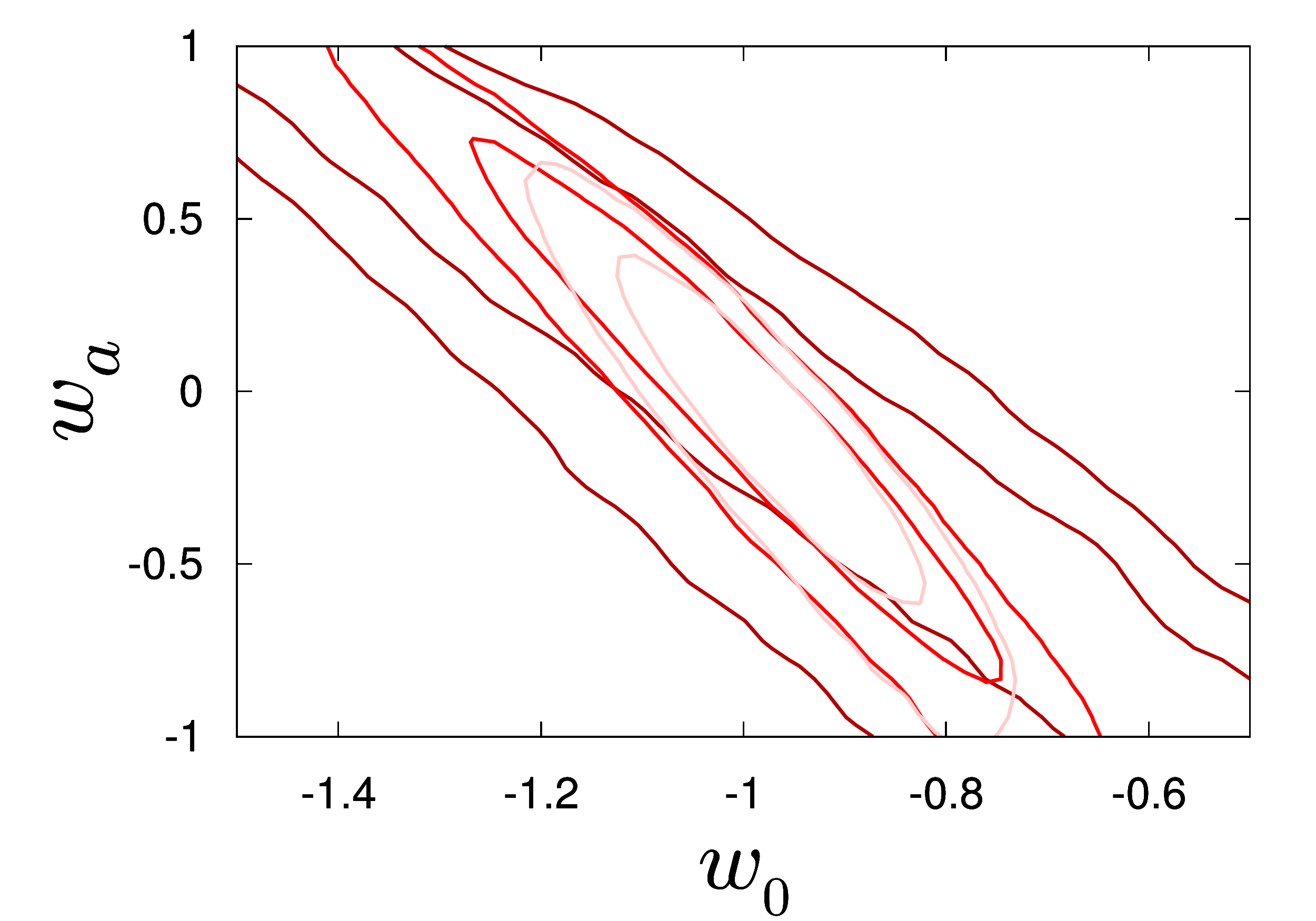}
\includegraphics[height=.48\textwidth,angle=270]{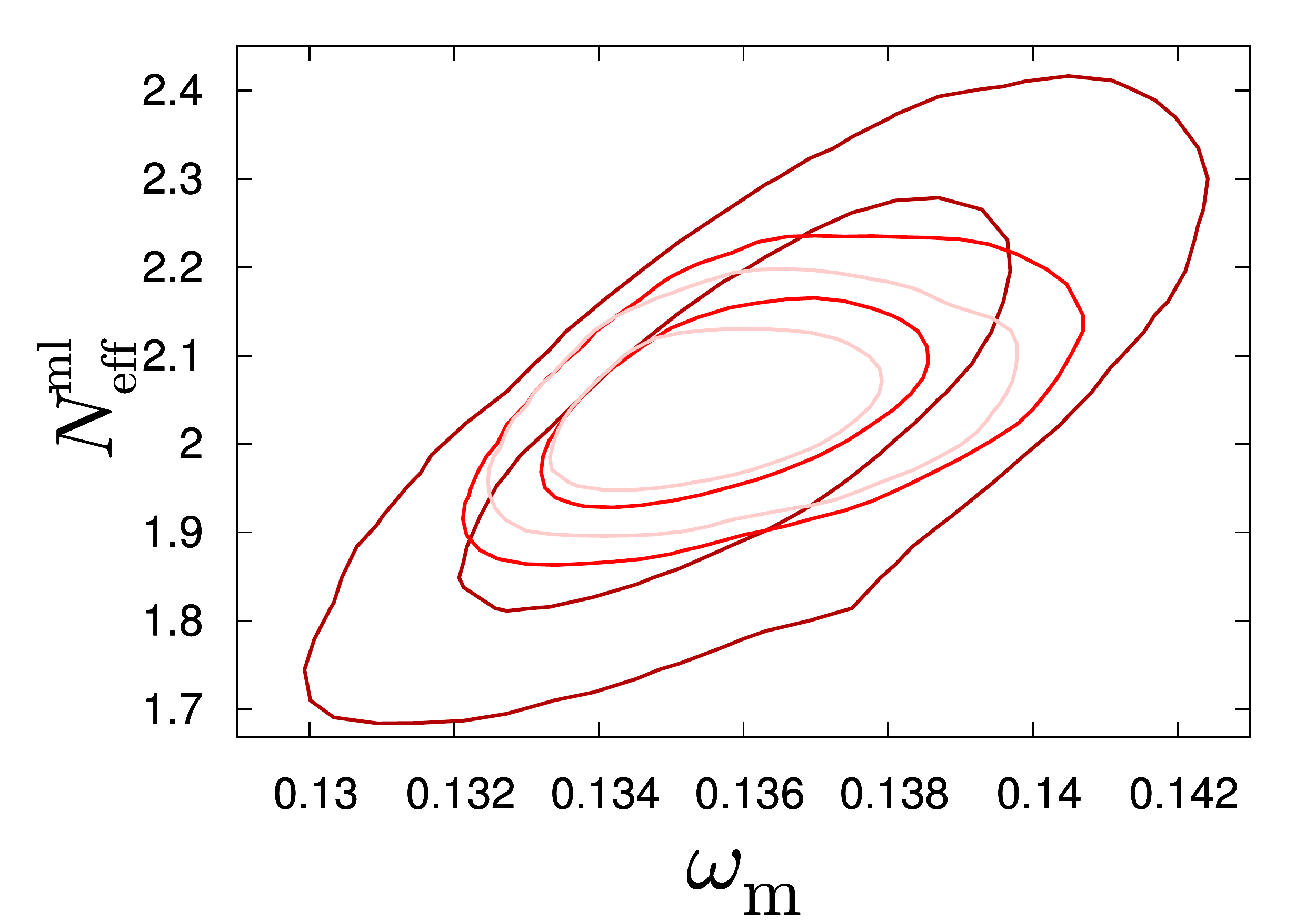}
\caption{Marginalised joint two-dimensional  68\% and 95\% credible contours from the CMB+clusters data set.  The default redshift and mass binning configuration for the cluster data is
$N_{\rm bin}=10$ (light red), but we also show the results for $N_{\rm bin}= 1$ bin (dark red) and 2  (red).}
\label{fig:w0wabin}
\end{figure}

Some parameter sensitivities continue to improve beyond $N_z=10$ and one might therefore argue for using a very large number of redshift bins. However, for $N_z=10$, the narrowest bin typically has a width of order $\Delta z = 0.03$; pushing much beyond $N_z=10$ may cause our results to lose their robustness against redshift uncertainty (see  section~\ref{sec:sigmaz}).   Furthermore, when cluster data are used in conjunction with angular power spectra from cosmic shear and/or galaxy clustering, the gain in going beyond $N_z=10$ is substantially reduced.  Henceforth, we shall adopt $N_z = N_\mathrm{m}  \equiv N_{\rm bin} = 10$.
Figure~\ref{fig:w0wabin} shows the marginalised joint two-dimensional  posteriors in the $(w_0,w_a)$- and  the $(N_{\rm eff}^{\rm ml},\omega_\mathrm{m})$-subspace
from CMB+clusters for this configuration, as well as for $N_{\rm bin}=1,2$.

\subsection{Probes of the expansion history versus probes of the power spectrum}\label{sec:expansionvspk}

One of the advantages of the cluster mass function (with redshift binning) is that it is highly sensitive to those parameters that govern the linear growth function and hence (in the case of standard gravity) the expansion history of the universe.  This makes redshift-binned cluster measurements powerful for constraining dark energy parameters, as well as for establishing the reduced matter density $\Omega_\mathrm{m} \equiv \omega_\mathrm{m}/h^2$.  Furthermore, because the normalisation of the cluster abundance is directly sensitive to the physical matter density $\omega_\mathrm{m}$,  the Hubble parameter $h$ can also be very effectively constrained.

\begin{table*}[t]
  \caption{Posterior standard deviations for the parameters $\omega_{\rm m}$, $h$, $\sum m_\nu$, $N^{\rm ml}_{\rm eff}$, $w_0$, and $w_a$ derived from various combinations of data sets.  Here, ``c'' denotes {\sc Planck} CMB data, ``g'' galaxy auto-spectrum (11 redshift bins), ``s'' shear auto-spectrum (2 bins), ``x'' shear-galaxy cross-correlation, and ``cl'' the cluster data (10 redshift bins, 10 mass bins). The table also shows the posterior standard deviation of $w_{\mathrm p}$ defined in section~\ref{sec:FoM} and the $w_{\mathrm p}$-$w_a$ figure-of-merit (FoM), defined in equation~(\ref{eq:fom}) as $(\sigma(w_{\mathrm p})\sigma(w_a))^{-1}$.\label{tab:errors}}
\begin{center}
{\footnotesize
  \hspace*{0.0cm}\begin{tabular}
  {lcccccccc} \hline \hline
  Data & $10^3 \times \sigma(\omega_{\rm m})$ & $100 \times \sigma(h)$ & $\sigma(\sum m_\nu)/$eV & $\sigma(N^{\rm ml}_{\rm eff})$ & $\sigma(w_0)$ & $\sigma(w_{\mathrm p})$ & $\sigma(w_a)$ & FoM$/10^3$ \\       \hline
  csgx & $1.3$ & $0.69$ & $0.023$ & $0.074$ & $0.13$ & $0.011$ & $0.19$ & $0.48$\\
  ccl & $1.6$ & $0.80$ & $0.050$ & $0.061$ & $0.24$ & $0.036$ & $0.33$ & $0.084$ \\
  csgxcl & $0.50$ & $0.46$ & $0.015$ & $0.033$ & $0.11$ & $0.0096$ & $0.15$ & $0.69$ \\
cscl & $0.67$ & $0.70$ & $0.020$ & $0.043$ & $0.17$ & $0.015$ & $0.24$ & $0.28$\\
  \hline \hline
  \end{tabular}
  }
  \end{center}
\end{table*}

\begin{figure}[t]
\includegraphics[height=.48\textwidth,angle=270]{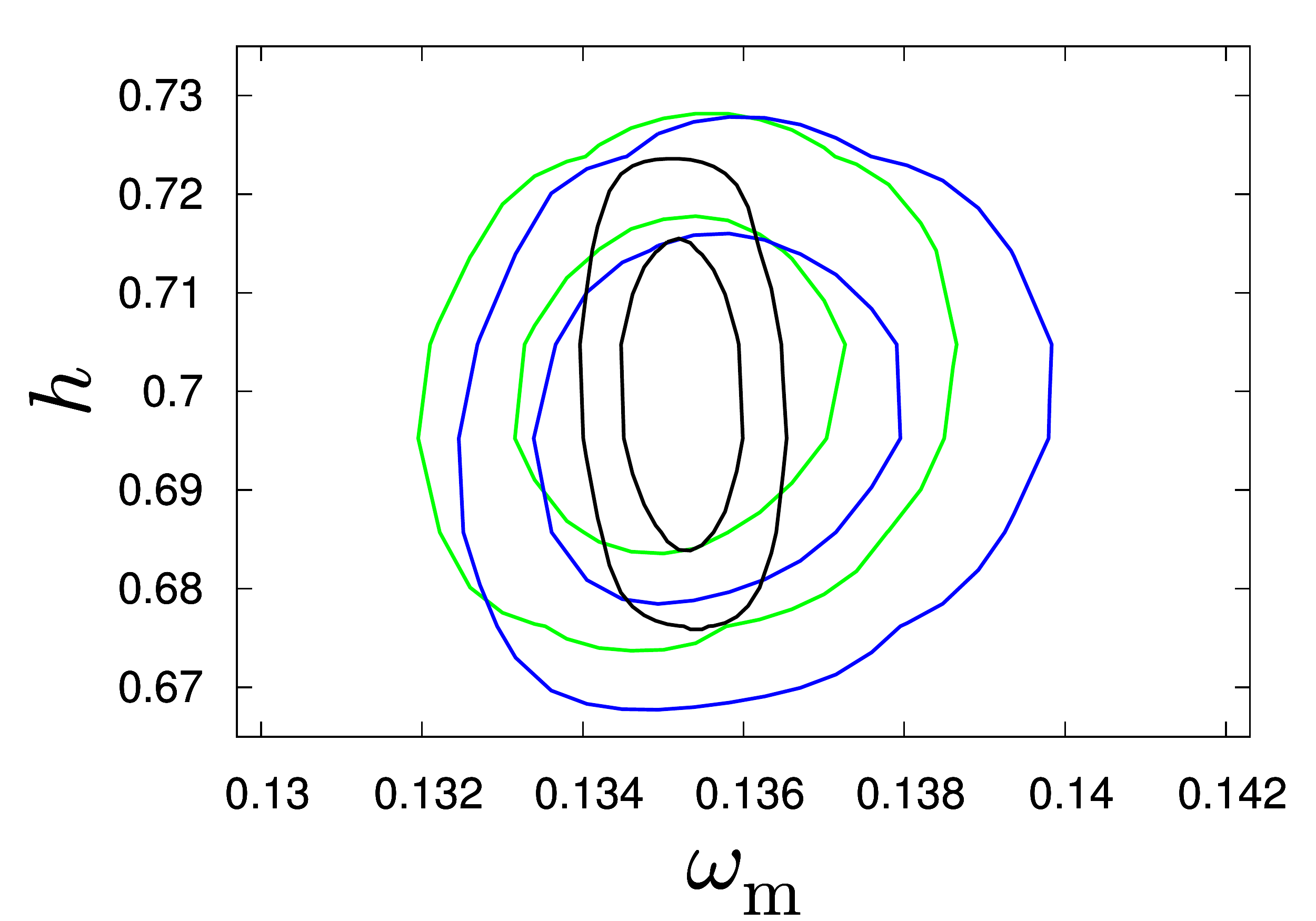}
\includegraphics[height=.48\textwidth,angle=270]{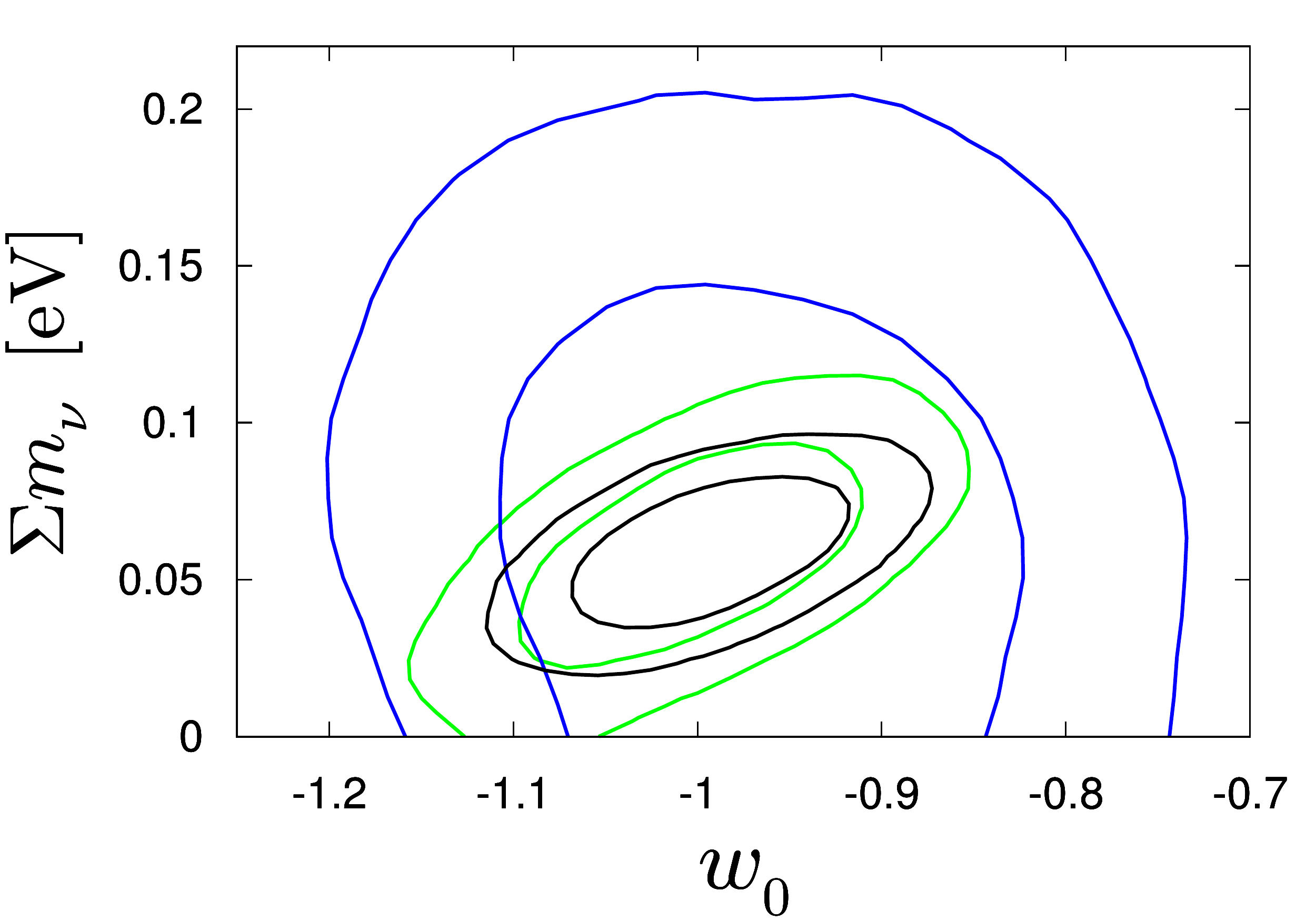} \\
\includegraphics[height=.48\textwidth,angle=270]{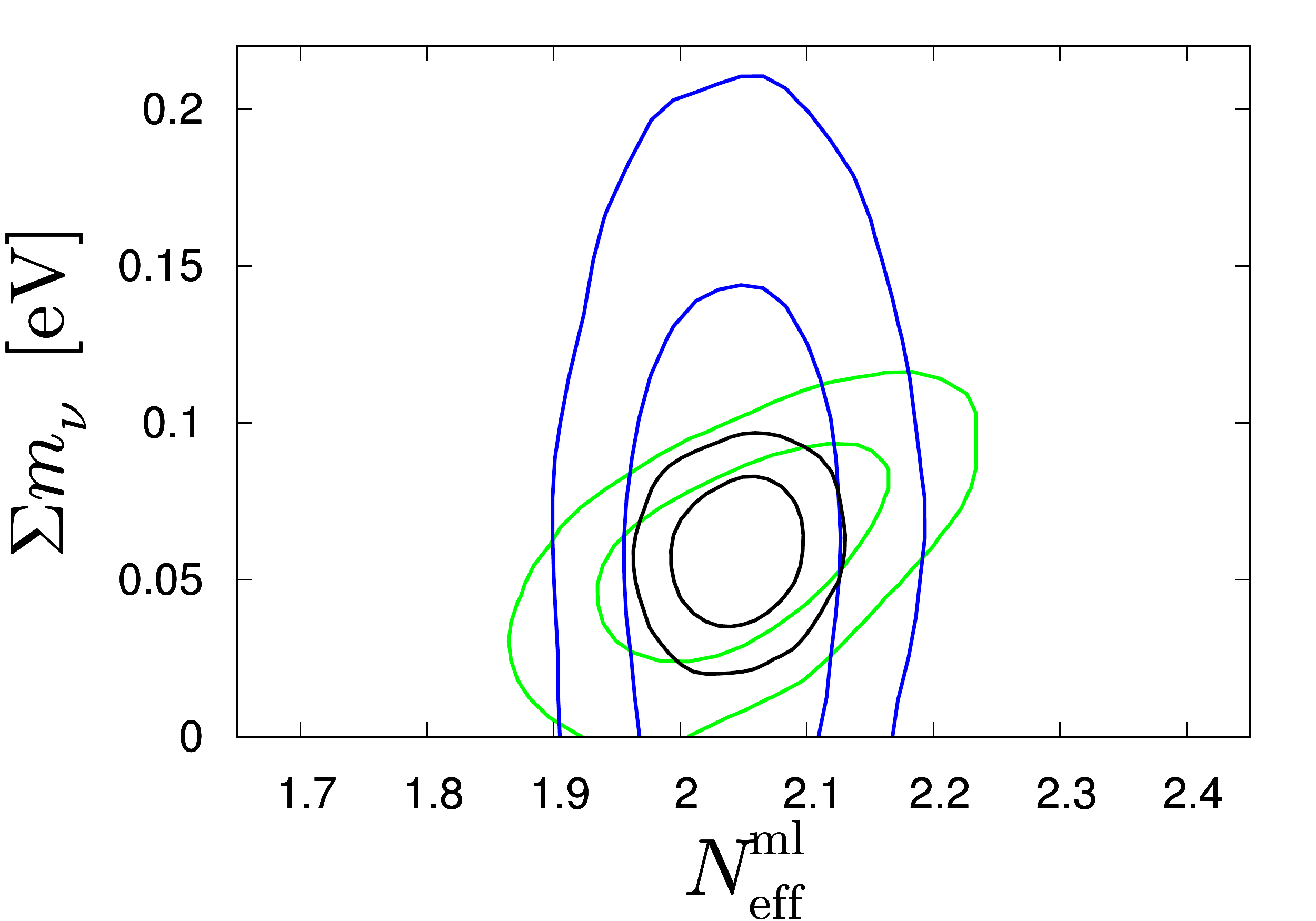}
\includegraphics[height=.48\textwidth,angle=270]{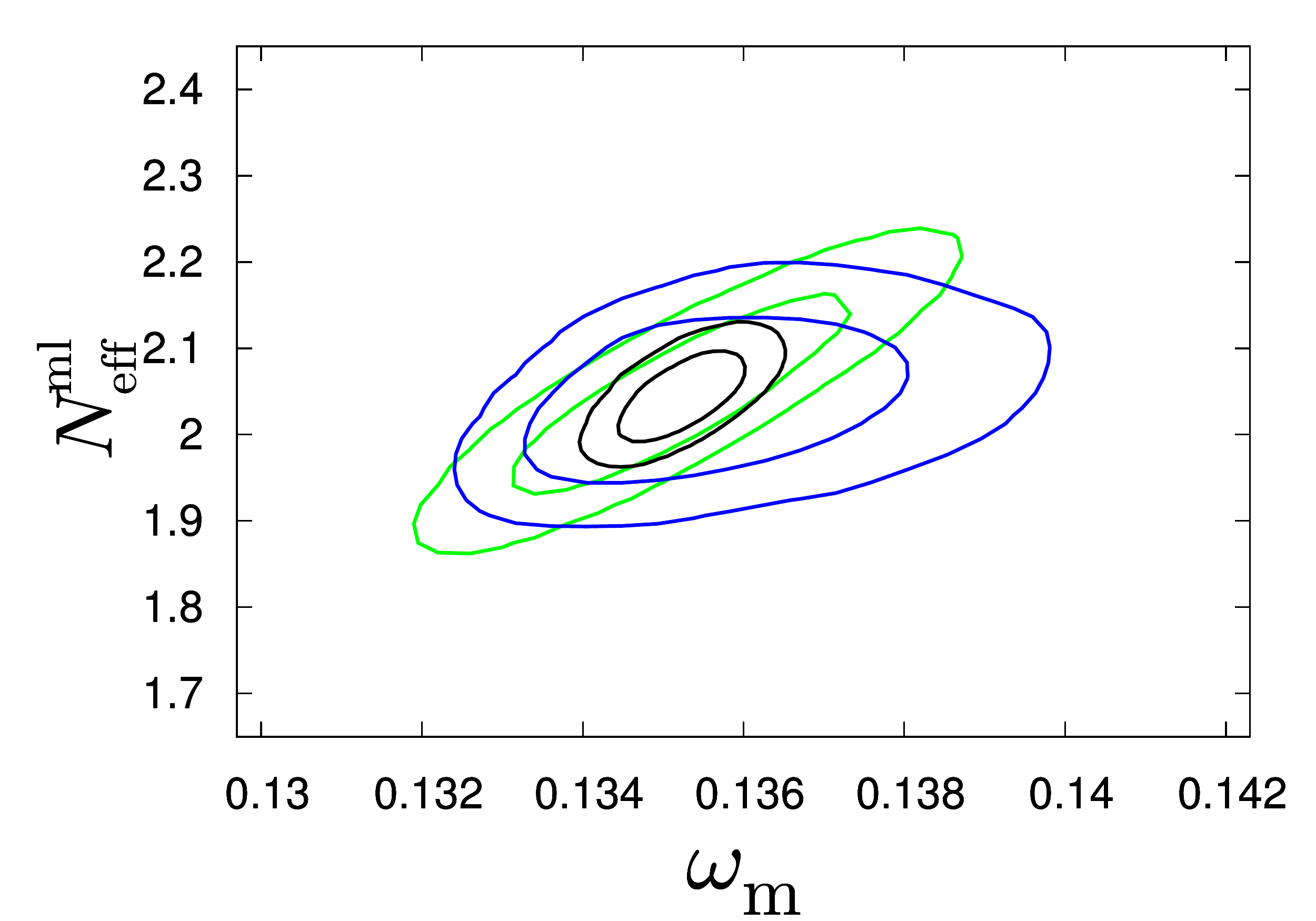}
\caption{Marginalised joint two-dimensional  68\% and 95\% credible contours from the CMB+clusters data set (``ccl'', blue), CMB+shear+galaxies (``csgx'', green),
and all data sets (``csgxcl'', black) for various parameters, using the default binning configuration of $N_{\rm bin}=10$ for the cluster data. \label{fig:contours}}
\end{figure}

The sum of neutrino masses $\sum m_\nu$ is significantly less well measured by clusters than by the shear and the galaxy power spectra.  This is because firstly, $\sum m_\nu$ plays a negligible role (compared with, e.g., dark energy parameters) in the redshift dependence of the late-time linear growth function.  Secondly, although the  shape of the cluster mass function is in principle also subject to a mass-dependent suppression due to neutrino free-streaming (e.g.,~\cite{Brandbyge:2010ge,Marulli:2011he}),  the actual range of cluster masses probed by a realistic cluster survey is very narrow (see figure~\ref{fig:illbin}), so that the suppression can be easily be mimicked by other effects such as an excess of relativistic energy density or simply a smaller initial fluctuation amplitude.

 Interestingly, a non-standard radiation content as parameterised by $N_{\rm eff}^{\rm ml}$, although has no direct effect on the late-time expansion or growth history, is quite well constrained by CMB+clusters.  This can be understood as follows: using CMB data alone, $N_{\rm eff}^{\rm ml}$ is strongly degenerate with
$\omega_\mathrm{m}$ and $h$.  However, because the cluster mass function is directly sensitive to $\omega_\mathrm{m}$ and $h$, it very effectively lifts any degeneracy of these parameters with $N_{\rm eff}^{\rm ml}$ when used in combination with CMB data.  As shown in the lower right panel of figure~\ref{fig:contours}, very little degeneracy remains between $N_{\rm eff}^{\rm ml}$ and $\omega_\mathrm{m}$ for the CMB+clusters data set.  A more telling illustration of how the binned cluster data removes the $(N_{\rm eff},\omega_\mathrm{m})$-degeneracy can be found in the right panel of figure~\ref{fig:w0wabin}:  Here, when only one redshift and mass bin is used, the cluster mass function is primarily sensitive to the fluctuation amplitude on small scales so that the $(N_{\rm eff},\omega_\mathrm{m})$-degeneracy persists in the CMB+clusters fit.
However, as soon as access to the linear growth function and some shape information become available through as little as  $N_z=N_\mathrm{m}=2$  bins, the degeneracy is partly broken because of the growth function's direct dependence on $\Omega_\mathrm{m}$ and of the normalisation's dependence on $\omega_\mathrm{m}$.

\subsection{Combining all data sets: constraints on neutrino parameters} \label{sec:other}

Perhaps the most noteworthy result of table~\ref{tab:errors} is that, while  CMB+shear+galaxies  (``csgx'') and CMB+clusters (``ccl'')  are well-suited to measuring different parameters and are hence in a sense complementary to each other, the combined usage of {\it all} data sets, i.e., the ``csgxcl'' combination, always leads to fairly significant enhancements in {\it all} parameter sensitivities.  This result can be understood from figure~\ref{fig:contours}, where it is clear that the `csgx'' and ``ccl''  datasets give rise to almost orthogonal parameter degeneracy directions.  In combination these data sets conspire to lift each other's degeneracies.

For neutrino parameters, it is interesting to note that while ``csgx'' and ``ccl'' return $\sigma(\sum m_\nu) = 0.023$~eV and 0.050~eV respectively,  in combination the sensitivity improves to $\sigma(\sum m_\nu) = 0.015$~eV.  This is nearly as good a sensitivity as was found earlier in~\cite{Hamann:2012fe} from the ``csgx'' data set, but for a much simpler 7-parameter cosmological model. This extraordinary sensitivity to $\sum m_\nu$ does not deteriorate much even if we exclude galaxy clustering from the analysis; as shown in table~\ref{tab:errors}, the ``cscl'' combination yields a similar $\sigma(\sum m_\nu)=0.020$~eV.  This is an especially reassuring result in view of the assumption of an exactly known linear galaxy bias we have adopted for the galaxy power spectrum, which some may deem unrealistic.
Thus, we again conclude that {\sc Euclid}, in combination with {\sc Planck} CMB data, will be able to probe neutrino masses at $3\sigma$~precision or better.  Likewise, the ``csgxcl'' data set is now sensitive to $N_{\rm eff}^{\rm ml}$ at $\sigma(N_{\rm eff}^{\rm ml})=0.033$ ($0.043$ for ``cscl''), meaning that for the first time the small deviation  of 0.046 from 3 in the fiducial~$N_{\rm eff}^{\rm ml}$ can be probed with $1\sigma$ precision.

\subsection{The dark energy figure-of-merit}\label{sec:FoM}

It is useful to quantify the constraining power of an observation (or a combination thereof) over a particular set of cosmological parameters in terms of a figure-of-merit (FoM). 
For the dark energy equation of state $w(a)$, one could define the FoM to be the inverse of the $N$-volume spanned by the error ellipsoid in the $N$-dimensional parameter space describing $w(a)$, such that the larger the volume the smaller the constraining power.  A na\"{\i}ve volume (or area) estimator in the 2-dimensional space of our parameterisation might be the product $\sigma(w_0)\sigma(w_a)$.
However, because $w_0$ and $w_a$ are strongly correlated, as is evident in figure~\ref{fig:w0wabin}, and the degree of correlation is {\it a priori} unknown, the product  
$\sigma(w_0) \sigma(w_a)$ will not only always overestimate the area of the ellipse, but do so also in a way that is strongly dependent on the degree of correlation between the two parameters.
For this reason, a more commonly adopted definition, used also in, e.g., the {\sc Euclid} Red Book~\cite{Laureijs:2011mu}, is
\begin{equation}
\label{eq:fom}
{\rm FoM} \equiv (\sigma(w_{\mathrm p})\sigma(w_a))^{-1},
\end{equation}
following from a parameterisation of the dark energy equation of state of the form $w(a) = w_{\mathrm p} + w_a (a_{\mathrm p}-a)$.
Here, the ``pivot'' scale factor $a_{\mathrm p}$ is chosen such that
$w_{\mathrm p}$ and $w_a$ are uncorrelated, i.e., the parameter directions are defined to align with the axes of the error ellipse.
Note that this parameterisation of $w(a)$ is entirely equivalent to the conventional $(w_0,w_a)$-parameterisation; the parameter spaces are related by a linear rotation, thus preserving the area of the error ellipse~\cite{Albrecht:2006um}.

Table \ref{tab:FoMexp}  contrasts the FoM computed as per definition~(\ref{eq:fom}) and the na\"{\i}ve estimate $(\sigma(w_0) \sigma(w_a))^{-1}$. For our default 10-parameter model, the
FoM of the ``csgxcl'' data combination is approximately 690, while the na\"{\i}ve approach underestimates the figure by about a factor of ten.
To facilitate comparison with other estimates in the literature, we also perform the same calculation for a reduced 7-parameter model,
\begin{equation}
\label{eq:reducedl}
\Theta^{\rm reduced} \equiv \left( \omega_{\rm b},\omega_{\rm dm},h,A_{\rm s},n_{\rm s},w_0,w_a \right),
\end{equation}
motivated in part by the model used in {\sc Euclid} Red Book~\cite{Laureijs:2011mu}.%
\footnote{The model adopted in reference~\cite{Laureijs:2011mu} has spatial curvature $\Omega_k$ also as a free parameter.  However, the degree of correlation between $\Omega_k$  and the dark energy parameters $w_0$ and $w_a$ is expected to be quite small for a {\sc Euclid}-like survey~\cite{Barenboim:2009ug}.  
We therefore adhere to our original assumption of spatial flatness,
but simply note that the FoM obtained in this work for the reduced parameter space~(\ref{eq:reducedl}) will in general be more optimistic than the value quoted in~\cite{Laureijs:2011mu}.\label{footnote:parameterspace}}
As expected, the smaller parameter space yields a significantly better FoM, about 1900, than our default 10-parameter model.  This figure is about 50\% lower than the official value of 4020 from the {\sc Euclid} Red Book~\cite{Laureijs:2011mu}, however, it climbs up to a higher value of about 5200 if we switch from MCMC to a Fisher matrix forecast such as that in~\cite{Laureijs:2011mu}.%
\footnote{Fisher matrix forecasts have a tendency to overestimate parameter sensitivities compared with MCMC analyses, a point also discussed in, e.g., \cite{Perotto:2006rj,Khedekar:2012sh,Wolz:2012sr}.}
Many factors could have contributed to the difference between our and the Euclid official FoM, from the assumed parameter space (see footnote~\ref{footnote:parameterspace})
to the survey parameters actually used in the analysis.  As the discrepancy is no more than 50\%, which could be interpreted as a reasonable compatibility, 
we shall not investigate its origin any further.  However, we stress that any FoM quoted for an observation or a set of observations is strongly dependent on the assumptions about the underlying cosmological parameter space, and therefore should always be taken {\it cum grano salis}.

\begin{table*}[t]
\caption{The dark energy figure-of-merit (FoM$/10^3$) computed using an MCMC forecast [MCMC($w_p,w_a$)] and 
a Fisher matrix forecast [Fisher($w_p,w_a$)]  for the two parameter spaces of equations~(\ref{eq:model}) and~(\ref{eq:reducedl}).  For
comparison, we also show the corresponding values obtained from a na\"{\i}ve estimate  [MCMC($w_0,w_a$)] as discussed in the main text.
See legend in table~\ref{tab:errors}. \label{tab:FoMexp}}
\begin{center}
{\footnotesize
  \hspace*{0.0cm}\begin{tabular}
  {llccc} \hline \hline
	Data & Parameter space & MCMC $(w_0,w_a)$ & MCMC $(w_{\mathrm p},w_a)$ & Fisher $(w_{\mathrm p},w_a)$  \\ \hline
	csgxcl & equation~\eqref{eq:model} ($\hat{c}_{\rm s}^2 = \infty$) & 0.060 & 0.69 &  \\
	csgxcl & equation~\eqref{eq:reducedl} & 0.30 & 1.9 & 5.2 \\
  \hline \hline
  \end{tabular}
  }
  \end{center}
\end{table*}

Lastly, we present in~table~\ref{tab:FoM} the FoMs for various fiducial models and data combinations.  Clearly, the FoM depends crucially on the data combination used to derive it; between the combinations ``ccl'' and ``csgxcl'', the difference in the FoMs is typically a factor of five to ten. When changing the fiducial cosmology, the trend is that a less negative equation of state at the present or in the past leads to a higher figure of merit. The dependence on the choice of fiducial values for the model parameters is however fairly weak; moving away from a $\Lambda$CDM fiducial cosmology ($w^{\rm fid}_0=-1,w^{\rm fid}_a=0$) induces no more than a 50\% variation in the FoM (provided the same number of parameters is varied).  We may therefore consider the FoM computed for $w^{\rm fid}_0=-1$ and $w^{\rm fid}_a=0$
as representative.

\begin{table*}[t]
  \caption{Posterior standard deviations for the parameters $w_0$, $w_{\mathrm p}$, and $w_a$ derived from the ``ccl'' and ``csgxcl'' data combinations
 (see legend in table~\ref{tab:errors}) assuming different fiducial values for the parameters.
 Also shown are the corresponding FoMs as per definition~(\ref{eq:fom}).  \label{tab:FoM}}
\begin{center}
{\footnotesize
  \hspace*{0.0cm}\begin{tabular}
  {lcccccc} \hline \hline
  Data & $w_0^{\rm fid}$ & $w_a^{\rm fid}$ & $\sigma(w_0)$ & $\sigma(w_{\mathrm p})$ & $\sigma(w_a)$ & FoM$/10^3$ \\       \hline
ccl & -1.00 & 0.00 & 0.24 & 0.036 & 0.33 & 0.084 \\
ccl & -0.83 & 0.00 & 0.18 & 0.027 & 0.26 & 0.14 \\
ccl & -1.17 & 0.00 & 0.31 & 0.044 & 0.39 & 0.058 \\
ccl & -1.00 & 0.35 & 0.17 & 0.029 & 0.25 & 0.14 \\
ccl & -1.00 & -0.35 & 0.28 & 0.041 & 0.38 & 0.064 \\
csgxcl & -1.00 & 0.00 & 0.11 & 0.0096 & 0.15 & 0.69 \\
csgxcl & -0.83 & 0.00 & 0.082 & 0.0087 & 0.12 & 0.96 \\
csgxcl & -1.17 & 0.00 & 0.13 & 0.011 & 0.18 & 0.51 \\
csgxcl & -1.00 & 0.35 & 0.075 & 0.0088 & 0.11 & 1.0 \\
csgxcl & -1.00 & -0.35 & 0.11 & 0.010 & 0.16 & 0.63 \\
  \hline \hline
  \end{tabular}
  }
  \end{center}
\end{table*}

\subsection{Dark energy sound speed and perturbations}
\label{sec:perturb}

In the analysis so far we have neglected the effect of dark energy perturbations.   We now introduce dark energy perturbations into the analysis as per the discussion in
section~\ref{sec:DEP}, and investigate the constraining power of a {\sc Euclid}-like survey on this aspect of dark energy.

\begin{figure}[t]
\center
\includegraphics[height=.48\textwidth,angle=270]{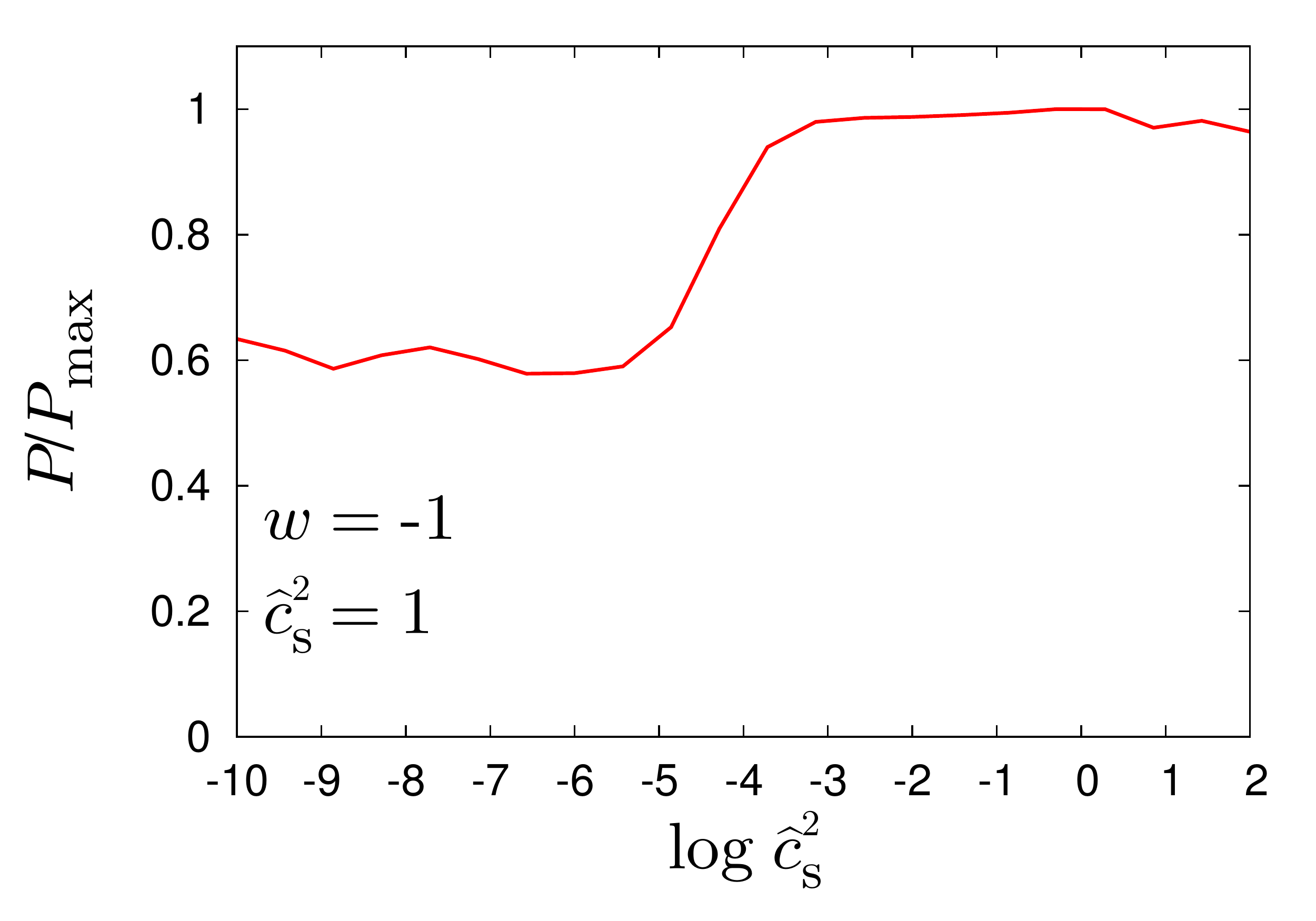}
\includegraphics[height=.48\textwidth,angle=270]{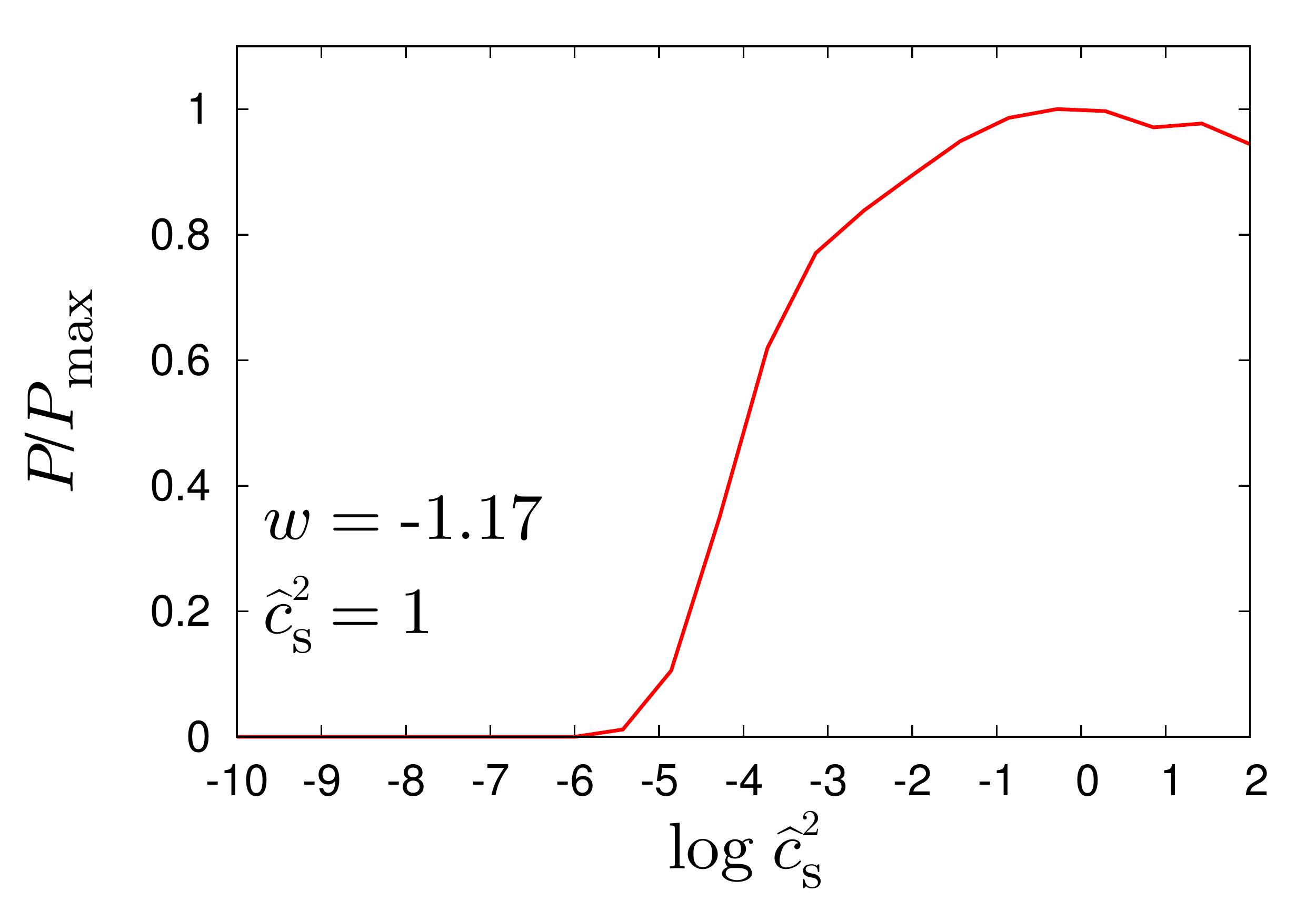}
\includegraphics[height=.48\textwidth,angle=270]{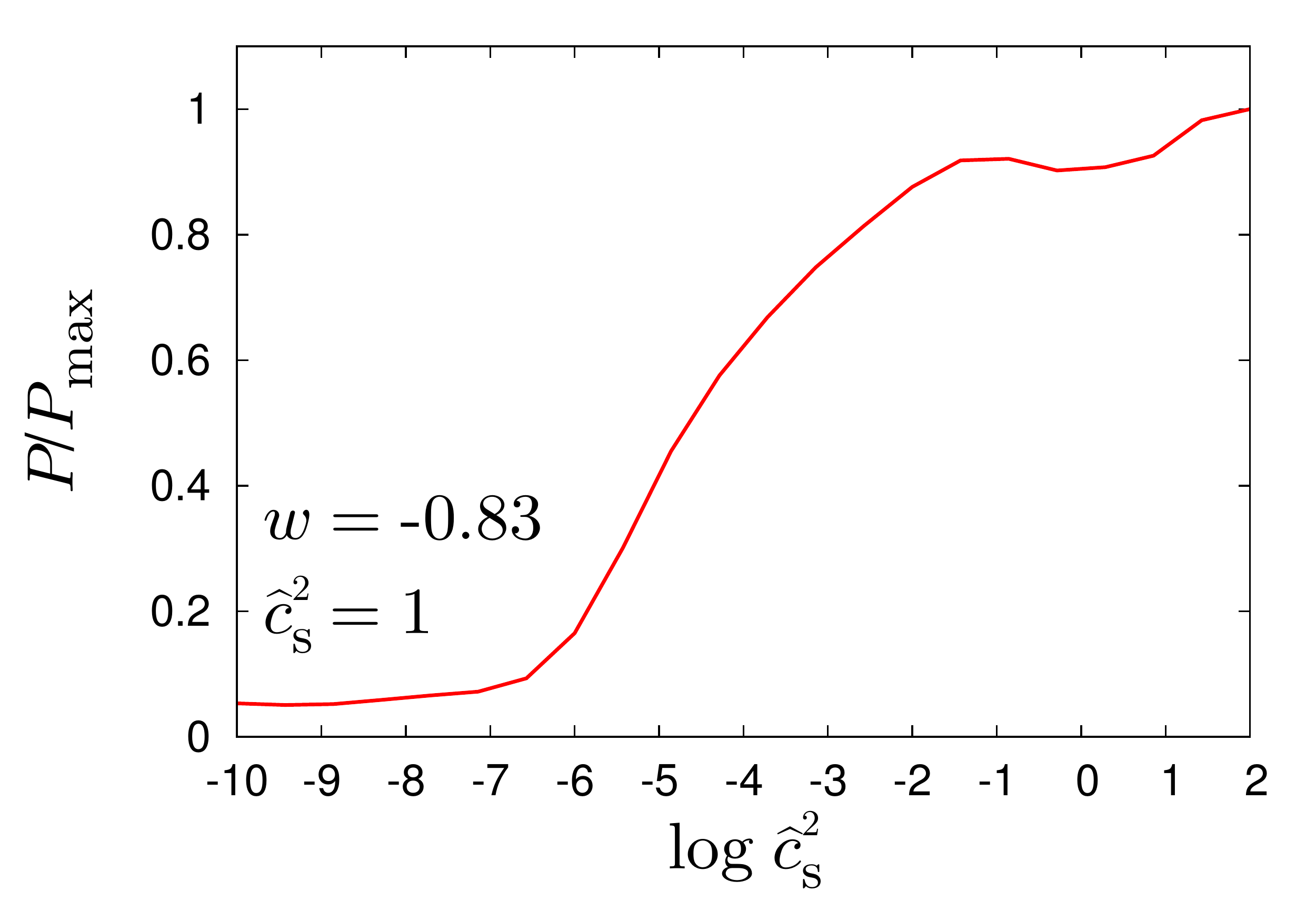}
\includegraphics[height=.48\textwidth,angle=270]{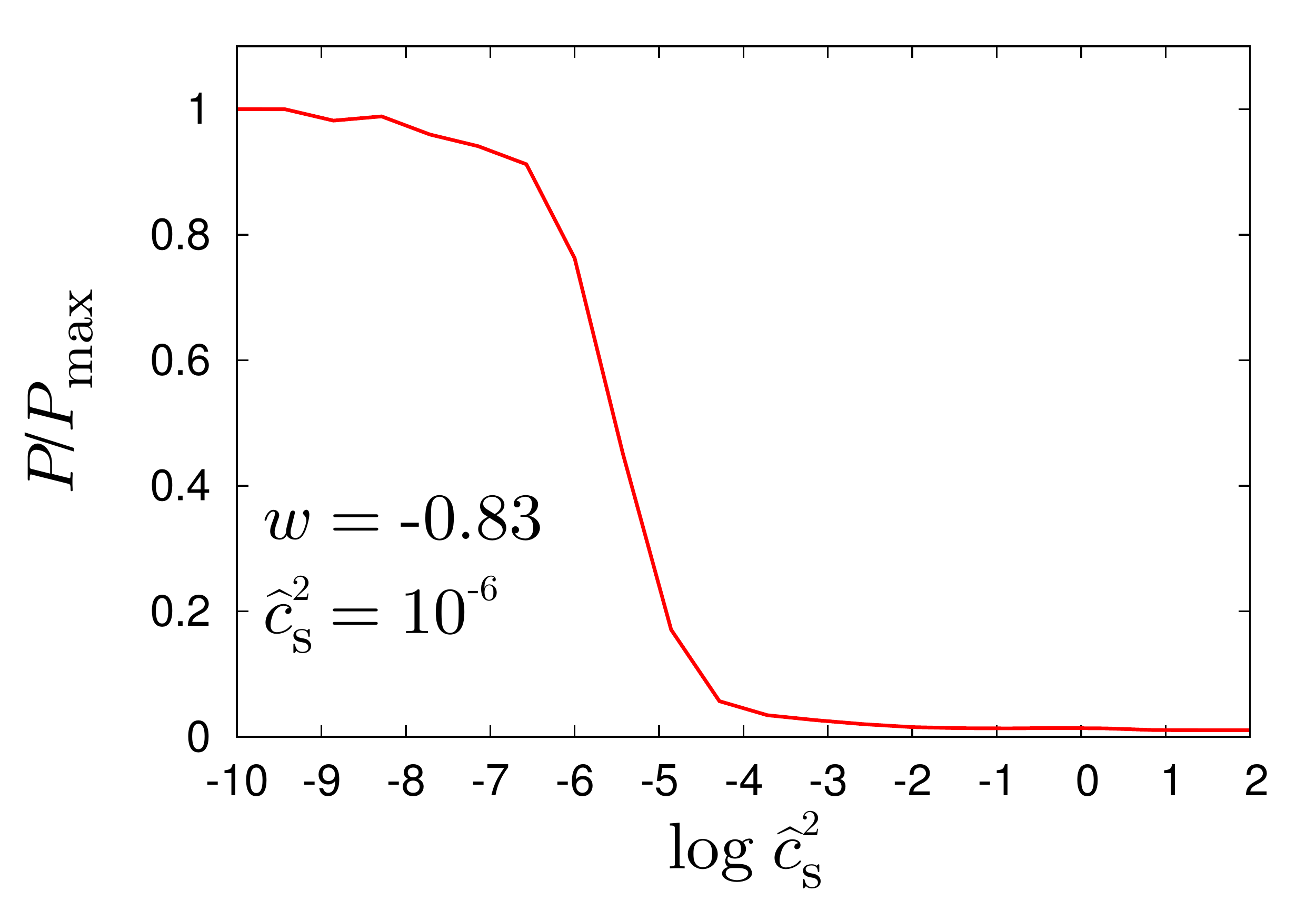}
\caption{Marginalised one-dimensional posterior probability density for $\log \hat{c}_\mathrm{s}^2$ for four different fiducial models (see labels in plots)
from a ``csgxcl'' fit.\label{fig:cscontours}}
\end{figure}

We consider four fiducial models differing in their  fiducial values of $(w_0,w_a,\hat{c}_\mathrm{s}^2)$:
\begin{itemize}

\item Model 1:  $(w_0,w_a) = (-1,0)$ and $\hat{c}_\mathrm{s}^ 2 =1$\footnote{The choice of $\hat{c}_\mathrm{s}^2$ for the $\Lambda$CDM model is immaterial, since by definition dark energy does not cluster in this model and $\delta_Q=0$ is automatically implemented in {\sc Camb}.}
(i.e., $\Lambda$CDM)

\item Model 2:  $(w_0,w_a)=(-1.17,0)$ and $\hat{c}_\mathrm{s}^2 =1$

\item Model 3:  $(w_0,w_a)=(-0.83,0)$ and $\hat{c}_\mathrm{s}^2 =1$

\item Model 4: $(w_0,w_a)=(-0.83,0)$ and $\hat{c}_\mathrm{s}^2 = 10^{-6}$

\end{itemize}
In all models, we fix $w_a = 0$ because of the pathological behaviour of dark energy perturbations when crossing the phantom divide $w = -1$~\cite{Vikman:2004dc,Hu:2004kh}. We present only results obtained from the ``csgxcl'' data sets, shown in table~\ref{tab:dep} and figure~\ref{fig:cscontours}.   All results have been obtained assuming a top-hat prior on $\log \hat{c}_\mathrm{s}^2$ of $\log \hat{c}_\mathrm{s}^2 \in [-10,2]$.
Note that the posteriors for $\log \hat{c}_\mathrm{s}^2$ in all models
show no clear peak structure so that $\log \hat{c}_\mathrm{s}^2$ can only be constrained from one side.  Instead of the posterior standard deviation,
we therefore quote in table~\ref{tab:dep} the appropriate one-sided limits.

\begin{table*}[t]
  \caption{Posterior 68\% (95\%) credible limits $\log \hat{c}_\mathrm{s}^2$ in various fiducial models derived from the combined ``csgxcl'' data set.
See legend in table~\ref{tab:errors}. 
         \label{tab:dep}}
\begin{center}
{\footnotesize
  \hspace*{0.0cm}\begin{tabular}
  {lcccccc} \hline \hline
  Data & $w_0$ & $w_a$ (fixed) & $\hat{c}_\mathrm{s}^2$ & $\log \hat{c}_\mathrm{s}^2$ & $\log \hat{c}_\mathrm{s}^2\; 68\%(95\%)$ C.I. \\       \hline
csgxcl & $-1.00$ & $0.00$ & $1$ & $0$ &  unconstrained \\
csgxcl & $-1.17$ & $0.00$ & $1$ & $0$ & $> -3.0(-4.1)$ \\
csgxcl & $-0.83$ & $0.00$ & $1$ & $0$ &   $>  -5.8(-)$\\
csgxcl & $-0.83$ & $0.00$ & $10^{-6}$ & $-6$   & $< -5.9(-3.5)$ \\
  \hline \hline
  \end{tabular}
  }
  \end{center}
\end{table*}

For the three models with $\hat{c}_\mathrm{s}^2 = 1$, the Jeans mass $M_\mathrm{J}$,  defined in equation~(\ref{eq:jeansmass}), is of order $10^{23}~{\rm M}_\odot$ at $z=0$ and
lies well above the maximum observed cluster mass $M_{\rm high}$.  This tells us immediately  that these models are undetectable by a {\sc Euclid}-like cluster survey.  However, as $\hat{c}_\mathrm{s}^2$ drops below $\sim 10^{-3}$ ($M_\mathrm{J} \sim 10^{17}~M_\odot$), the additional mass-dependence it produces on the cluster mass function begins to be visible to {\sc Euclid}, and we see a sharp decrease in the posterior probability for $\log \hat{c}_\mathrm{s}^2$ in figure~\ref{fig:cscontours}.  In the case of model 2, the corresponding posterior probabilities eventually drop to zero, {\color {red} thereby (was originally ``thus'')} allowing us to place a lower limit on $\hat{c}_\mathrm{s}^2$. The same trend can be seen also  for model 3, although in this case the posterior probability does not drop all the way to zero. A likely reason for this is that the main constraining power towards $\hat{c}_\mathrm{s}^2$ comes from the cluster survey and $w_0 > -1$ leads to fewer clusters compared to $w_0 < -1$, e.g., model 3 has about two thirds of the number clusters in model 2. This makes the effect of the dark energy perturbations less significant compared to the Poisson noise. No constraints on $\hat{c}_\mathrm{s}^2$ are available for model~1 (the $\Lambda$CDM model), because dark energy perturbations are generally suppressed by a factor of $1+w_0$ relative to dark matter perturbations and are hence practically nonexistent in the vicinity of $w_0=-1$. 

In the remaining model 4 with $\hat{c}_\mathrm{s}^2 = 10^{-6}$ the dark energy sound speed can be constrained from above. This is a consequence of $M_\mathrm{J}$ ($ \simeq 4 \times 10^{14} \, M_\odot$ at $z=0$) lying close to the detection threshold of the {\sc Euclid} cluster survey, i.e., the effect of the dark energy perturbations disappears when the sound speed is increased. Our studies show that with a lower detection threshold, so that the Jeans mass falls well within the cluster mass range probed by {\sc Euclid}, the transition from minimal dark energy clustering to full dark energy clustering can be very effectively observed with the help of cluster mass binning. This would allow the dark energy sound speed to be constrained from both sides. We emphasise however that our choice of the fiducial value of $w_0$ also plays a crucial role towards establishing the quoted limits, again because of the $1+w_0$ suppression suffered by the dark energy perturbations relative to the dark matter perturbations. When comparing the constraints obtained on model~3 and 4,  one should remember that the mass and redshift bins are assigned according to the fiducial model, so the signature of dark energy clustering at some value of $\hat{c}_\mathrm{s}^2$ appears different in the two models. The quantative constraints on $\log \hat{c}_\mathrm{s}^2$ obtained here should be taken {\it cum gran salis} as discussed in the next section.

\subsubsection{Modelling of the cluster mass funcion}\label{sec:modcmf}

The constraints obtained thus far on the dark energy sound speed are based on our particular modelling of the cluster mass function, namely, equation~(\ref{eq:sh}) and the prescriptions of appendix~\ref{sec:appendix}.  In this model, the cluster mass function is assumed to take the Sheth-Tormen form as a function of the {\it total cluster mass} 
$M=M_\mathrm{m}+M_Q$, i.e., including both masses of the nonrelativistic matter and the virialised dark energy.  Thus there are three ways in which the dark energy sound speed is propagated into the observable: (i) via the linear power spectrum into $\sigma_{\rm m}(M)$, (ii) the mass-dependent linear threshold density $\delta_c(M,z)$, and (iii) the virial mass of the cluster.

However, our model is by no means unique.  Other models exist and differ from ours essentially in the implementation of the above three points.  We emphasise that until  full-scale numerical simulations including clustering dark energy become available, it is not clear which of these models is the most correct.  For this reason, it is useful to test how strongly the constraints on $\hat{c}_{\rm s}^2$ depend on the modelling of the cluster mass function.  Here, we consider two variations:
\begin{enumerate}
\item The model of, e.g.,~\cite{Creminelli:2009mu,Batista:2013oca} assumes the cluster mass function to follow the Sheth-Tormen form as a function of the {\it nonrelativistic matter mass}, i.e.,
\begin{equation}
	\frac{\mathrm{d}n_{\rm ST}}{\mathrm{d}M_\mathrm{m}}(M,z)=-\sqrt{\frac{2a}{\pi}}A\left[1\!+\!\left(\! \frac{a\delta_c^2}{\sigma_\mathrm{m}^2}\right)^{-p}\right]\frac{\bar\rho_\mathrm{m}}{M_\mathrm{m}^2}\frac{\delta_c}{\sigma_\mathrm{m}}\! \left(\frac{\mathrm{d}\log \sigma_\mathrm{m}}{\mathrm{d}\log M_\mathrm{m}}\!-\!\frac{\mathrm{d}\log \delta_c}{\mathrm{d}\log M_\mathrm{m}}\right)\exp\! \left[\! -a\frac{\delta_c^2}{2\sigma_\mathrm{m}^2}\right],
 \label{eq:mcmf}
\end{equation}
where $\sigma_\mathrm{m}^2(M_m,z)$ is the variance of the linear matter density field smoothed on a comoving length scale $X_{\rm sm} \equiv a^{-1} [3 M_\mathrm{m}/(4 \pi \bar{\rho}_\mathrm{m})]^{1/3}$.   The observed cluster mass $M$ is identified with the total virial mass of the cluster $M_{\rm vir} = M_\mathrm{m}+M_Q(\tau_{\rm vir})$, 
which in the arbitrary $\hat{c}_{\rm s}^2$ case can be established using our interpolation method described in appendix~\ref{sec:appendix}.

This definition of the cluster mass function will in general result in weaker constraints on the dark energy sound speed compared with our default model, 
because the effect of the virialised dark energy $M_Q$ on the observable is factored in only through the selection function of the survey.
We explicitly test this definition for the fiducial model~4 ($w_0=-0.83, w_a=0, \hat{c}_{\rm s}^2 = 10^{-6}$), and find that the sensitivity to $\log \hat{c}_{\rm s}^2$ degrades significantly (see table~\ref{tab:modcs}).

\item Another possibility is to completely ignore dark energy perturbations in the nonlinear modelling of the collapse (i.e., the spherical collapse) of the cluster.  
In this case dark energy perturbations affect the observable quantity only through their effects on the linear matter power spectrum. This leads to even fewer features in the cluster distribution and hence even weaker constraints on the dark energy sound speed.  Again, table~\ref{tab:modcs} shows that, assuming the fiducial model~4, the constraints on $\log \hat{c}_{\rm s}^2$ degrade.

\end{enumerate}

\begin{table*}[t]
  \caption{Posterior 68\% (95\%) credible limits $\log \hat{c}_\mathrm{s}^2$ for various definitions of the cluster mass function (CMF) derived from the ``csgxcl'' data combination.
See legend in table~\ref{tab:errors}.
         \label{tab:modcs}}
\begin{center}
{\footnotesize
  \hspace*{0.0cm}\begin{tabular}
  {llccccc} \hline \hline
  Data & CMF & $w_0$ & $w_a$ (fixed) & $\hat{c}_\mathrm{s}^2$ & $\log \hat{c}_\mathrm{s}^2$ & $\log \hat{c}_\mathrm{s}^2\; 68\%(95\%)$ C.I. \\       \hline
csgxcl & equation~\eqref{eq:sh} & $-0.83$ & $0.00$ & $10^{-6}$ & $-6$ & $< -5.9(-3.5)$ \\
csgxcl & equation~\eqref{eq:mcmf} & $-0.83$ & $0.00$ & $10^{-6}$ & $-6$ & $< -1.4(1.4)$ \\
csgxcl & $P^\mathrm{lin}_\mathrm{m}(k,z)$ only & $-0.83$ & $0.00$ & $10^{-6}$ & $-6$ & $< -2.5(0.12)$ \\
  \hline \hline
  \end{tabular}
  }
  \end{center}
\end{table*}

In summary, any constraint on the dark energy sound speed derived from cluster measurements is strongly dependent on the modelling of the cluster mass function.  To this end, 
a full-scale numerical simulation is mandatory to establish the definitive model, in order for any quoted constraint on $\hat{c}_{\rm s}^2$ to be meaningful.

\section{Conclusions}\label{sec:conc}

In this paper we have considered the constraining power of a {\sc Euclid}-like galaxy survey on cosmological parameters in conjunction with {\sc Planck} CMB data. This study is  an extension of our previous investigation in~\cite{Hamann:2012fe}, in that we have included in the present analysis mock data from the {\sc Euclid} cluster survey in addition to the angular cosmic shear and galaxy power spectra expected from the photometric redshift survey, and we have expanded the parameter space to encompass also dynamical dark energy as well as the possibility of (small-scale) dark energy perturbations.

We find that the different combinations of data sets, CMB+clusters and CMB+shear+galaxies, give comparable sensitivities for parameters that affect only the late-time growth and expansion history of the universe, i.e., those parameters that determine the dynamical dark energy equation of state and the Hubble parameter.  The constraints for CMB+clusters depend chiefly on our adoption of redshift binning for the observed clusters, which allows us in particular to probe the transition from matter to dark energy domination.
 Neutrino masses, on the other hand, are not particularly well-constrained by CMB+clusters ($\sigma(\sum m_\nu) = 0.050$~eV in a 10-parameter model), clearly because they do not play a major role in the  overall linear growth of matter density perturbations.  Importantly, however, the degeneracy directions of CMB+clusters and CMB+shear+galaxies are largely orthogonal.  This means that even though neither data set performs particularly impressively for any one cosmological parameter, in combination they help to lift each other's degeneracies.  The
sensitivities to $\sum m_\nu$ from CMB+shear+galaxies+clusters is, for example, $\sigma(\sum m_\nu) = 0.015$~eV in a 10-parameter model, which is almost as good as that obtained previously in~\cite{Hamann:2012fe} from CMB+shear+galaxies for a much simpler 7-parameter $\Lambda$CDM model.  Thus, we can conclude again that a {\sc Euclid}-like survey has the potential to measure neutrino masses at $4 \sigma$ precision or more.

For the dark energy parameters, we find that the combination of CMB+shear+galaxies+ clusters results in a dark energy figure-of-merit (FoM), defined in this work as $(\sigma(w_{\mathrm p}) \sigma(w_a))^{-1}$, of~690 for a $\Lambda$CDM fiducial cosmology, with variations of up to 50\% for fiducial cosmologies in which $w_0 \neq -1$ and $w_a \neq 0$.   We emphasise that this number has been derived for a 10-parameter cosmological model.  Were we to adopt instead a simpler 7-parameter model in which $\sum m_\nu$, $N_{\rm eff}$ and $z_{\rm re}$ are fixed, then the FoM would climb up to 1900 (and to 5200 if a Fisher matrix analysis, instead of MCMC, was used), a value that is comparable to the official estimate of 4020 from the {\sc Euclid} Red Book~\cite{Laureijs:2011mu}.

Finally, we investigate the detectability of  dark energy perturbations, parameterised in terms of a non-adiabatic fluid sound speed $\hat{c}_\mathrm{s}^2$.  Along the way  we also introduce a model of the cluster mass function that incorporates the effects of  $\hat{c}_\mathrm{s}^2$ based on solving and interpolating the spherical top-hat collapse in the known limits of  $\hat{c}_\mathrm{s}^2 \to  \infty$ (homogeneous dark energy) and $\hat{c}_\mathrm{s}^2 =0$ (dark energy comoving with nonrelativistic matter).   We find that for values of the dark energy sound speed whereby the associated Jeans mass lies within the mass detection range of the cluster survey, dark energy perturbations imprint a distinct step-like signature in the observed cluster mass function. With the help of cluster mass binning, this signature makes these models distinguishable from those in which the Jeans mass lies well outside (both below and above) the detection range. The models tested in this paper have associated Jeans masses either well above the mass range probed or close to the detection threshold, and we show that these can be distinguished at $2\sigma$, as long as the fiducial value of $w_0$ deviates from $-1$ by as much as is presently allowed by observations.  

We emphasise however that constraints on the dark energy sound speed from cluster measurements depend strongly on the modelling of the cluster mass function, with our default model being a very optimistic one. The very large sensitivity range clearly illustrates the enormous uncertainties in the current state of cluster mass function modelling for clustering dark energy cosmologies.  The need for 
full-scale numerical simulations including dark energy perturbations cannot be overstated if future observations are to be interpretable in these contexts.

{\bf Note added}: As this work was in its final stage of completion,  we learnt of the investigation of~\cite{Cerbolini:2013uya} which considered the {\sc Euclid} cluster survey's sensitivities to neutrino parameters.  While a full comparison is difficult because of the generally different assumptions about the survey parameters and the model parameter space, where the assumptions do to some extent coincide the two analyses appear to be compatible.

\section*{Acknowledgements}

We acknowledge computing resources from the Danish Center for Scientific Computing (DCSC). OEB acknowledges support from the Stellar Astrophysics Centre at Aarhus University. We thank Ronaldo Batista for discussions on the modelling of the cluster mass function in the context of clustering dark energy.

\appendix

\section{Interpolating the spherical collapse between the two limits of \texorpdfstring{$\hat{c}_\mathrm{s}^2$}{cs2}}\label{sec:appendix}

The spherical top-hat collapse model is exactly defined only in the limits $\hat{c}_\mathrm{s}^2 \to \infty$ and $\hat{c}_\mathrm{s}^2 =0$.  In the first case, the dark energy component is non-clustering.  The dark matter and baryon components alone suffer gravitational collapse, so that the overdense region preserves its top-hat density profile throughout the collapse, with a comoving radius $X$  given by
\begin{equation}
\label{eq:sphericalaaa}
\frac{\ddot{X}}{X} + {\cal H}
\frac{\dot{X}}{X} = - 4 \pi G a^2 \bar{\rho}_\mathrm{m} \delta_\mathrm{m},
\end{equation}
where
\begin{equation}
\delta_\mathrm{m} (\tau) =  [1 + \delta_\mathrm{m}(\tau_i)] \left[ \frac{X(\tau_i)}{X(\tau)} \right] ^3-1
\end{equation}
follows from conservation of the total mass of nonrelativistic matter $M_\mathrm{m}$ in the top-hat region.  The virial radius $R_{\rm vir}$, defined as the physical radius of the top-hat at the moment the collapsing region fulfils the virial theorem, is as usual one half of physical radius at turn-around, and the virial mass $M_{\rm vir}$ is identical to $M_\mathrm{m}$, which we also equate with the mass of the cluster $M$.

In the second case, an exactly vanishing non-adiabatic dark energy sound speed means that, like nonrelativistic matter, dark energy density perturbations also evolve identically on all scales.  This again leads to the preservation of the top-hat density profile and hence the conservation of $M_\mathrm{m}$ in the region defined by the comoving radius $X$, now determined by
\begin{equation}
\label{eq:spherical}
\frac{\ddot{X}}{X} + {\cal H} \frac{\dot{X}}{X} = - 4 \pi G a^2 [\bar{\rho}_\mathrm{m} \delta_\mathrm{m} + \bar{\rho}_Q  \delta_Q].
\end{equation}
A conservation law can likewise be written down for the clustered dark energy component
\begin{equation}
\label{eq:conserve}
\dot{\rho}_Q + 3 \left( {\cal H} + \frac{\dot{X}}{X} \right) (\rho_Q + \bar{P}_Q) = 0,
\end{equation}
where $\rho_Q$ denotes the dark energy density in the top-hat region.  We assume that at any one time the clustered dark energy contributes a mass~\cite{Creminelli:2009mu}
\begin{equation}
M_Q(\tau)  \equiv \frac{4 \pi}{3} \bar{\rho}_Q (\tau) \delta_Q(\tau) [a(\tau) X(\tau)]^3
\end{equation}
to the total mass of the system.  This clustered dark energy takes part in virialisation, defined here as the instant the system satisfies the condition $\mathrm{d}^2I/\mathrm{d}t^2 =0$,
where $I  \equiv (2/5) (M_\mathrm{m}+M_Q) (a X)^2$ is the top-hat's moment of inertia, and $t$ is the cosmic time ($\mathrm{d}t = a \mathrm{d}\tau$).  The physical radius of the top-hat at this instant is the virial radius $R_{\rm vir}  \equiv a X(\tau_{\rm vir})$, and the total mass the virial mass $M_{\rm vir} \equiv M_\mathrm{m} + M_Q(\tau_{\rm vir})$.
Linearised forms of equations~(\ref{eq:sphericalaaa}) to~(\ref{eq:conserve}) can be found
in references~\cite{Creminelli:2009mu,Basse:2010qp}.

Between the two limits, any finite nonvanishing $\hat{c}_\mathrm{s}^2$ necessarily causes the overdense region to evolve away from the top-hat configuration and to become ill-defined.  The absence of strict conservation laws in this intermediate regime also renders the system not readily soluble.
Nonetheless, the transition between the two known limits have been investigated using a quasi-linear approach in~\cite{Basse:2010qp,Basse:2012wd}, where
it was found that, for a fixed~$\hat{c}_\mathrm{s}^2$ and at a given collapse redshift~$z$, the transition generically results in a step-like feature in the linear threshold density $\delta_c$, as well as in the quantities $\Delta_{\rm vir} \equiv 3M_\mathrm{m}/4\pi\bar{\rho}_\mathrm{m}(\tau_{\rm vir})R^3_{\rm vir}$ and \mbox{$\eta_{\rm vir} \equiv M_Q(\tau_{\rm vir})/M_\mathrm{m}$}, as $M_\mathrm{m}$ is varied from $M_\mathrm{m} \ll M_\mathrm{J}$ to $M_\mathrm{m} \gg M_\mathrm{J}$, with
\begin{eqnarray}
\label{eq:jeansmass}
M_\mathrm{J}(a) & \equiv& \frac{4 \pi}{3} \bar{\rho}_\mathrm{m}(a) \left[\frac{a \lambda_\mathrm{J} (a)}{2} \right]^3 \nonumber \\
& = &  9.7 \times 10^{23} \  c_\mathrm{s}^3 \ \Omega_\mathrm{m} a^{-3} \left(\Omega_\mathrm{m} a^{-3} + \Omega_Q e^{-3 \int_0^a \mathrm{d}a'  [1+w(a')]/a'} \right)^{-3/2}
\ h^{-1} M_\odot
\end{eqnarray}
denoting the Jeans mass, and $\lambda_\mathrm{J} \equiv 2 \pi \hat{c}_\mathrm{s}/{\cal H}$  the corresponding comoving Jeans length.

To incorporate this feature in the cluster mass function for our MCMC analysis, our strategy is as follows.

\begin{enumerate}

\item For each given set of cosmological parameters, we solve the spherical collapse model for $\delta_c$, $\Delta_{\rm vir}$, and $\eta_{\rm vir}$ in the two limits of $\hat{c}_\mathrm{s}^2$
using equations~(\ref{eq:sphericalaaa}) to~(\ref{eq:conserve}) as functions of the collapse redshift~$z$.

\item At each redshift, we interpolate the two limits using the formulae
\begin{eqnarray}
\label{eq:fitting}
&& \delta_c(M_\mathrm{m}) = \Delta(\delta_c^\infty,\delta_c^0)\tanh \left[A_1 \left(\log M_\mathrm{m} - \log\frac{M_\mathrm{J}}{B_1}\right) \right] + \Sigma(\delta_c^\infty,\delta_c^0), \\
&& \Delta_{\rm vir}(M_\mathrm{m}) = \Delta(\Delta_{\rm vir}^\infty ,\Delta_{\rm vir}^0) \tanh \left[A_2 \left(\log M_\mathrm{m} - \log\frac{M_\mathrm{J}}{B_2}\right) \right] + \Sigma(\Delta_{\rm vir}^\infty , \Delta_{\rm vir}^0), \\
&& \eta_{\rm vir} (M_\mathrm{m}) = \Delta(\eta_{\rm vir}^\infty,\eta_{\rm vir}^0) \tanh \left[A_3 \left(\log M_\mathrm{m} \!-\! \log\frac{M_\mathrm{J}}{B_3}\right) \right] \!+\!\Sigma(\eta_{\rm vir}^\infty,\eta_{\rm vir}^0),
\end{eqnarray}
where $x^\infty$ denotes the value of $x=\delta_c,\Delta_{\rm vir},\eta_{\rm vir}$ in the $\hat{c}_\mathrm{s}^2 \to \infty$ limit, $x^0$ the $\hat{c}_\mathrm{s}^2=0$ limit, and
$\Delta(x^\infty,x^0) \equiv(x^\infty-x^0)/2$ and $\Sigma(x^\infty,x^0)\equiv(x^\infty+x^0)/2$ represent their difference and sum respectively.
 The parameters $A_{1,2,3}$ and $B_{1,2,3}$ are fitting coefficients, adjusted to fit respectively the sharpness and location of the transition.
  At $z=0$ and for $(w_0,w_a)=(-0.8,0)$, we find that $A_{1,2,3} \simeq 1.4$, and $B_{1,2,3} \simeq 20$ reproduce the quasi-linear results of~\cite{Basse:2010qp,Basse:2012wd} in the region immediately below the Jeans mass quite well.  For simplicity, we adopt these parameter values for all redshifts and cosmological models.

\item Lastly, we identify the virial mass with the cluster mass, i.e., $M_{\rm vir} \equiv M$, and construct the functions $\delta_c(M,z)$ and $R_{\rm vir}(M,z)$ according to
\begin{eqnarray}
&& \delta_c(M,z) \equiv \delta_c(M_{\rm vir}(M_\mathrm{m}),z) \equiv \delta_c(M_\mathrm{m},z),\\
&& R_{\rm vir}(M,z) \equiv R_{\rm vir}(M_{\rm vir}(M_\mathrm{m}),z) \equiv R_{\rm vir}(M_\mathrm{m},z),
\end{eqnarray}
which are used in sections~\ref{sec:theory} and~\ref{sec:thr} determine the cluster mass function and the mass detection threshold, respectively.

\end{enumerate}

\bibliographystyle{utcaps}

\bibliography{refs}

\end{document}